\documentclass[prb,amsmath,amssymb,floatfix,nofootinbib,superscriptaddress,reprint,twocolumn]{revtex4-1}
\usepackage{graphicx}
\usepackage{hyperref}
\usepackage{algorithmic}
\usepackage{algorithm}
\usepackage{appendix}
\usepackage{rotating}
\usepackage{txfonts}
\usepackage{array}
\usepackage{bm}
\usepackage{dcolumn}
\usepackage{amsmath}
\usepackage{color}
\usepackage{physics}
\usepackage{braket}
\usepackage[caption=false]{subfig}

\usepackage{xcolor}
\hypersetup{
    colorlinks,
    linkcolor={red!50!black},
    citecolor={blue!50!black},
    urlcolor={blue!80!black}
}
\newcommand{\crop}[1]{\ensuremath{\hat{a}_{#1}^\dagger}}
\newcommand{\anop}[1]{\ensuremath{\hat{a}_{#1}}}

\newcommand{\change}[1]{{#1}} 

\begin{document}
\title{Localized Quantum Chemistry on Quantum Computers}
\date{\today}

\author{Matthew \surname{Otten}}
\email{mjotten@hrl.com}
\affiliation{HRL Laboratories, LLC, 3011 Malibu Canyon Road,
Malibu, CA 90265}
\author{Matthew R.\ \surname{Hermes}}
\affiliation{Department of Chemistry, Pritzker School of Molecular Engineering, James Franck Institute, Chicago Center for Theoretical Chemistry, University of Chicago, Chicago, IL 60637, USA.}
\author{Riddhish \surname{Pandharkar}}
\affiliation{Department of Chemistry, Pritzker School of Molecular Engineering, James Franck Institute, Chicago Center for Theoretical Chemistry, University of Chicago, Chicago, IL 60637, USA.}
\author{Yuri Alexeev}
\affiliation{Computational Science Division, Argonne National Laboratory, Lemont, IL 60439, USA}
\author{Stephen K.\ \surname{Gray}}
\email{gray@anl.gov}
\affiliation{Center for Nanoscale Materials,
Argonne National Laboratory, Lemont, IL 60439, USA}
\author{Laura \surname{Gagliardi}}
\email{lgagliardi@uchicago.edu}
\affiliation{Department of Chemistry, Pritzker School of Molecular Engineering, James Franck Institute, Chicago Center for Theoretical Chemistry, University of Chicago, Chicago, IL 60637; Argonne National Laboratory, Lemont, IL 60439, USA.}
\begin{abstract}
Quantum chemistry calculations of large, strongly correlated systems are typically limited by the computation cost that scales exponentially with the size of the system. Quantum algorithms, designed specifically for quantum computers, can alleviate this, but the resources required are still too large for today's quantum devices. Here we present a quantum algorithm that combines a localization of multireference wave functions of chemical systems with quantum phase estimation (QPE) and variational unitary coupled cluster singles  and doubles (UCCSD) 
to compute their ground state
energy. Our algorithm, termed ``local active space unitary coupled cluster'' (LAS-UCC), scales linearly with system size for certain geometries, providing a polynomial reduction in the total number of gates compared with QPE, while
providing accuracy above that  of the variational quantum eigensolver  using the UCCSD ansatz and also above that  of the classical local active space self-consistent field. The accuracy of LAS-UCC is demonstrated by dissociating (H$_2$)$_2$ into two H$_2$ molecules and by breaking the two double bonds in \emph{trans}-butadiene and resources estimates are provided for linear chains of up to 20 H$_2$ molecules.
\end{abstract}
\maketitle

\section{Introduction \label{sec:intro}}

Chemical systems with many close-lying electronic states or, more generally, strongly correlated electrons pose a significant challenge for modern electronic structure theories in computational quantum chemistry~\cite{lowdin1958correlation,sherrill2007bond,krylov2007breaking,stein2014seniority,gaggioli2019beyond}. When transition metals or heavier elements are involved,  degenerate and 
nearly degenerate electronic states are common, and single-reference electronic structure methods such as Kohn--Sham density functional theory often fail~\cite{neese2009prediction,jacob2012spin,yu2016perspective}. In these situations one has to use multireference methods to generate multiconfigurational wave functions and accurately describe these near degeneracies~\cite{szalay2012multiconfiguration,park2020multireference,cramer2006theoretical}.

Scientists also want to compute properties of large chemical systems or solids with accurate quantum chemistry methods, in spite of steep computational requirements. One way to achieve such  computations is to use fragmentation methods. Many variations of fragmentation methods exist~\cite{gordon2012fragmentation, collins2015energy, raghavachari2015accurate, fedorov2011geometry}, but the common feature is that a large molecular system is divided into fragments and quantum-mechanical calculations are performed on the fragments. An especially important case is the application of fragmentation methods to multireference wave functions because of the exponential explosion of the computational cost with respect to the size of the active space of electronic configurations.

In the complete active space self-consistent field (CASSCF) method~\cite{Roos1980}, all the electronic configurations that can be formed for a given number of active electrons distributed in a given number of 
active orbitals are included in the wave function. Thus,  the wave function scales exponentially with the number of active electrons and orbitals, and  the method has only limited application to chemically relevant systems. If one wants to study systems containing, for example, several transition metals~\cite{LiManni2021,hallmen2019toward,sharma2019multiple,hogue2018spin,malrieu2014magnetic}, the active site of a protein~\cite{Levine2020}, or extended organic chains in their ground and excited states~\cite{Levine2020,Sharma2019b}, more affordable multireference methods have to be developed. This is one of the major challenges of modern electronic structure theory.

Reducing  the computational cost  of CASSCF or other multiconfiguration self-consistent  field calculations is pursued both in the development of new well-motivated theoretical approximations and in the application of new developments in computational hardware~\cite{Hohenstein2015,Snyder2016}. On the theoretical side, one strategy is to identify subspaces of the CAS that can be treated on different footings~\cite{olsen1988determinant,malmqvist1990restricted} or interact with one another only weakly~\cite{ma2011generalized,Ivanic2003,Parker2014,Nishio2019,Kathir2020}. 
\change{The localized active-space self-consistent field (LASSCF) method~\cite{Hermes2019,Pandharkar2019a,Hermes2020,pandharkar2021localized}, also known as the cluster mean-field (cMF) method,~\cite{Jimenez-Hoyos2015} is an example of such a strategy}. LASSCF is designed for applications in which electrons are strongly  correlated in different weakly interacting physical regions of a molecule and approximates the strongly  correlated part of the wave function as a single antisymmetrized product of subspace wave functions. The computational cost of LASSCF is a linear function of the number of such unentangled subspaces. 

Some of the authors have recently shown that LASSCF accurately reproduces the CASSCF spin-state energy gaps of bimetallic compounds and the simultaneous dissociation of two double bonds in bisdiazene at a significantly reduced cost~\cite{Pandharkar2019a,Hermes2020}. However, LASSCF fails to recover any electron correlation between fragments, for example in the \textit{cis-trans} isomerization of stilbene and similar systems~\cite{pandharkar2021localized}.
\change{Moreover, methods to restore the missing correlation variationally~\cite{Abraham2020c}, perturbatively~\cite{Jimenez-Hoyos2015,Papastathopoulos-Katsaros2021}, or \emph{via} the coupled-cluster (CC) approach~\cite{Wang2020b} on classical computers must usually enumerate a general many-body basis for each fragment. That is, they inherit the complications of multireference perturbation and CC theory~\cite{Lyakh2012,park2020multireference} over traditional single-reference perturbative or truncated coupled-cluster (CC) corrections based on second quantization~\cite{Moller1934,ShavittBartlett_textbook}.} 

Recently, the development of quantum computers has led to an increased interest in novel quantum algorithms, especially for computational quantum chemistry, which is widely seen as a potential ``killer app'' of quantum computers~\cite{cao2019quantum,head2020quantum,alan_2020_review}. \change{The quantum phase estimation (QPE) quantum algorithm~\cite{lloyd1996universal}  can potentially offer exponential speedups when large fault-tolerant quantum computers are available~\cite{kitaev1995quantum,abrams1999quantum},
under the assumption that an initial state with non-negligible overlap can be prepared~\cite{kitaev2002classical,o2021electronic}.
Additionally, the variational unitary coupled cluster (UCC) requires
only a polynomial number of gates to represent on a quantum computer, whereas representing the same ansatz classically has no known polynomial solution~\cite{romero2018strategies,peruzzo2014variational}.}
For the noisy, intermediate-scale quantum (NISQ)~\cite{preskill2018quantum} devices that we have today, these algorithms are not tenable, since they require coherence times far beyond what is available. Variational algorithms, such as the variational quantum eigensolver (VQE)~\cite{peruzzo2014variational}, have been used to perform calculations of the ground state energy of small molecules, with limited accuracy, on NISQ devices~\cite{wecker2015progress,mcclean2017hybrid,kandala2017hardware}. Quantum algorithms that have less stringent requirements compared with full QPE, and at the same time accuracy beyond that demonstrated by variational algorithms such as VQE, will be required to productively use the progressively larger and higher-quality quantum devices as they become available in the next few years.

In this paper we describe a framework for such quantum algorithms, inspired by classical LASSCF. The wave function within a fragment is solved by using one method (e.g., QPE), and correlation between fragments is encoded variationally by using an ansatz that entangles the fragments. This approach goes beyond what can be achieved with classical fragment methods, such as LASSCF, by providing additional correlation between fragments, while significantly reducing the total computational time (estimated via the number of gates) compared with full
QPE. 

\section{Theory \label{sec:theory}}

\subsection{Multireference Methods with Exponential Scaling}
We seek to find the ground state of the second-quantized molecular Hamiltonian for a given number of $M$ electrons,
\begin{equation}\label{qchem_ham}
  \hat{H} = h^p_q \crop{p}\anop{q} + \frac{1}{4} h^{pr}_{qs} \crop{p}\crop{r}\anop{s}\anop{q},
\end{equation}
where $\crop{p}$ ($\anop{p}$) creates (annihilates) an electron in spin orbital $p$; $h^p_q$ and $h^{pr}_{qs}$ are the one- and antisymmetrized two-electron Hamiltonian matrix elements, respectively; and repeated internal indices are summed. Generally, for $N$ spin orbitals, $\hat{H}$ has a
sparse-matrix representation in a space of size $O\binom{N}{M}$
and has $O(N^4)$ elements. Full-configuration interaction (FCI) determines the exact energy within a given one-electron basis set
(the FCI energy) at exponential cost. Methods such as CASSCF (and its restricted \cite{olsen1988determinant,malmqvist1990restricted} and generalized \cite{fleig2001generalized,ma2011generalized} active space approximations) or
selected configuration interaction~\cite{li2020accurate,li2018fast}, can go beyond FCI in system size, maintaining comparable accuracy, but still scale exponentially. The density matrix renormalization group~\cite{olivares2015ab,knecht2015new,kurashige2011second,marti2011density} and coupled cluster methods~\cite{ShavittBartlett_textbook} can scale polynomially 
but introduce (sometimes
uncontrollable) approximation errors. Here we briefly
describe the LASSCF algorithm~\cite{Hermes2019,Hermes2020}, which 
will serve as the basis for our fragment-based quantum algorithms. 

\subsection{LASSCF \label{sec:theory_lasscf}}

In LASSCF, the wave function of a molecule is approximated as
\begin{equation}
    \ket{{\rm LAS}} = \bigwedge_K \ket{\Psi_K} \wedge \ket{\Phi}, \label{eq:las_wfn}
\end{equation}
where $\ket{\Psi_K}$ is a general many-body wave function describing $M_K$
electrons occupying $N_K$ active orbitals of the $K$th ``fragment'' or ``active
subspace,'' $\ket{\Phi}$ is a single determinant spanning the complement of the
complete active space, and the wedge operator (``$\wedge$'') implies an
antisymmetrized product. 

In the variational\cite{Hermes2020} implementation of LASSCF, this wave function
is obtained by minimizing the LAS energy, 
\begin{equation}
    E_{\rm{LAS}} = \braket{{\rm LAS}|\hat{H}|{\rm LAS}}, \label{eq:elas}
\end{equation} 
with respect to all orbital rotations and configuration interaction (CI) vectors defining $\ket{\rm{LAS}}$. This is accomplished by introducing a unitary operator (see the Supporting Information of Ref.\ \onlinecite{Hermes2020}) that is parameterized in terms of all nonredundant transformations of the orbitals and CI vectors,
\begin{equation}
    \ket{{\rm LAS}} \to \hat{U}_{\mathrm{orb}}\prod_K \hat{U}_{\mathrm{CI},K}\ket{{\rm LAS}}, \label{eq:ulas}
\end{equation}
where
\change{
\begin{eqnarray}
    \hat{U}_{\mathrm{orb}} &=& \exp{x^k_l\left(\crop{k}\anop{l}-\crop{l}\anop{k}\right)}, \label{eq:uorb} \\
    \hat{U}_{\mathrm{CI},K} &=& \exp{x_{\vec{k}}\left(\ket{\vec{k}}\bra{\Psi_K}-\ket{\Psi_K}\bra{\vec{k}}\right)}, \label{eq:ucik}
\end{eqnarray}
}
where $k,l$ index individual spin orbitals in two different subspaces (including the
inactive and virtual subspaces outside of the CAS) and where $\ket{\vec{k}}$ is a determinant or configuration state function. First and second derivatives of Eq.\ (\ref{eq:elas}) with respect to the generator amplitudes ($x^{k}_{l}$ and $x_{\vec{k}}$) are obtained by using the Baker--Campbell--Hausdorff (BCH) expansion, and the energy is minimized by repeated applications of the preconditioned conjugate gradient (PCG) method~\cite{Bernhardsson1999,Stalring2001}.

The orbital unitary operator, $\hat{U}_{\mathrm{orb}}$, corresponds to the
UCC correlator truncated after the first (``singles'')
term: 
\begin{eqnarray}
    \hat{U}_{\rm UCC} &\equiv& \exp{\hat{T}_{\rm UCC}}, \label{eq:ucc} \\ 
    \hat{T}_{\rm UCC} &\equiv& x^k_l \left(\crop{k}\anop{l}-\mathrm{h.c.}\right)
    + \frac{1}{4}x^{km}_{ln} \left(\crop{k}\crop{m}\anop{n}\anop{l} - \mathrm{h.c.}\right) + \ldots. \label{eq:tcc}
\end{eqnarray}

\change{The use of the more general cluster operator, Eq.\ (\ref{eq:tcc}), in place of the orbital rotation unitary operator, Eq.\ (\ref{eq:uorb}), corresponds to a multireference unitary coupled cluster method~\cite{Hoffmann1988} built on top of a $\ket{\rm LAS}$ reference wave function. Such a method is expected to more flexible than LASSCF itself, in that doubles and higher-order cluster amplitudes could encode electron correlation and entanglement between active subspaces. This would require the reference
wave function, $\ket{\rm LAS}$, to be updated by explicit exponentiation of 
the general cluster operator, Eq.\
(\ref{eq:tcc}), after each execution of the PCG algorithm. On classical
computer hardware, however, this is not an efficient way to extend LASSCF.} 

\subsection{LAS Methods on Quantum Computers \label{sec:theory_qlas}}
Here we describe an algorithm for molecular calculations that goes beyond the limited accuracy of standard VQE~\cite{kandala2017hardware,mcclean2017hybrid},
while having dramatically reduced computational complexity compared with QPE [see Methods section]. The
algorithm exploits the structure of the molecule by separating it into
coupled fragments, as is done in the classical algorithm, LASSCF. The quantum
algorithm, however, goes beyond classical LASSCF by providing some degree of
entanglement between the fragments.

The algorithm  begins by segmenting the orbital active space of a given molecule into distinct fragments defined by non-overlapping orbital subspaces, as in classical LASSCF. For instance, orthogonalized atomic orbitals generated by using the meta-L\"{o}wdin method\cite{Sun2014} can be sorted into localized fragments and then projected onto a guess for the CAS of a given molecule to produce localized active orbitals. We construct an effective Hamiltonian that omits non-mean-field interfragment interactions, resulting in a sum of local fragment Hamiltonians,
\begin{equation}
    \hat{H}_{{\rm eff}} = \sum_K^{n_f} \Bigg( \tilde{h}^{k_1}_{k_2} \crop{k_1}\anop{k_2} + \frac{1}{4} h^{k_1k_3}_{k_2k_4} \crop{k_1}\crop{k_3}\anop{k_4}\anop{k_2}\Bigg),
\end{equation}
where $k_1,k_2,\ldots$ index distinct active orbitals of the $K$th fragment and where
\begin{eqnarray}
\tilde{h}^{k_1}_{k_2} &=& h^{k_1}_{k_2} + h^{k_1i}_{k_2i}
+ \sum_{L\neq K} h^{k_1l_1}_{k_2l_2} 
\gamma^{l_1}_{l_2},
\label{eq:h1eff}
\end{eqnarray}
where $i$ and $l_n$ index respectively inactive orbitals [i.e., those defining $\ket{\Phi}$ in Eq.\ (\ref{eq:las_wfn})] and active orbitals of the $L$th fragment and where $\gamma^{l_1}_{l_2}$ is a \change{one-electron reduced} density matrix element for spin orbitals $l_1$ and $l_2$\change{,
\begin{equation}
    \gamma^{l_1}_{l_2} \equiv \braket{{\rm LAS}|\crop{l_1}\anop{l_2}|{\rm LAS}} = \braket{\Psi_L|\crop{l_1}\anop{l_2}|\Psi_L}.
\end{equation}}

Given a set of localized active orbitals that minimize the LASSCF energy, if the density matrices in Eq.\ (\ref{eq:h1eff}) are obtained from a classical LASSCF calculation on the same system, then the QPE algorithm applied to $\hat{H}_{{\rm eff}}$ generates the active-space part of the LASSCF wave function, $\ket{{\rm QLAS}} = \bigwedge_K \ket{\Psi_K}$, on the quantum computer. The same result is achieved if density matrices are obtained self-consistently from the QPE evaluation. If the density matrices are obtained in some other way, for instance from $\ket{{\rm HF}}$, then an approximation to the LASSCF wave function is obtained.

\change{The QPE step provides the initial $\ket{{\rm QLAS}}$ for each fragment step by repeating the measurement of the phase until it is consistent with the phase representing the ground state energy, 
which collapses the system into the ground state wavefunction. This introduces some overhead, as each fragment will need to be in the ground state to continue to the next step. Furthermore,
a full QPE solve, estimating the ground state energy, must be performed initially to provide a comparison value.}

A sequence of UCC with singles and doubles (UCCSD) circuits, with variable parameters, is then
applied across $m$ fragments each (which we term $m$-local), 
leading to the LAS-UCC wave function,

\change{
\begin{equation}
\ket{{\rm QLAS}(\bm{x})} \rightarrow \prod_\zeta \hat{U}_{{\rm UCCSD},\zeta}(\bm{x}) \ket{{\rm QLAS}}, \label{eq:mlocal}
\end{equation}
where $\hat{U}_{{\rm UCCSD},\zeta}(\bm{x})$ is the UCCSD ansatz including
only creation/annihilation operators within the $m$ fragments
that it spans, $\zeta$ is a list of fragment indices of size $m$, and $\bm{x}$ are the associated singles and doubles
cluster amplitudes.}
\change{The factorization of Eq.\ (\ref{eq:tcc}) implied by Eq.\ (\ref{eq:mlocal}) is based on the intuition that physically adjacent active subspaces are likely to be more strongly entangled to one another than subspaces on opposite ends of a large molecule.}
The parameters of the UCCSD circuit are varied to minimize the total energy of the full system, as in VQE \change{[see also Methods section]:
\begin{equation}\label{eq:vqe}
  E = \min_{\bm{x}} \braket{ {\rm QLAS}(\bm{x}) | \hat{H} |{\rm QLAS}(\bm{x})  }.
\end{equation}
}

\begin{figure}
    \centering
    \includegraphics[width=0.95\linewidth]{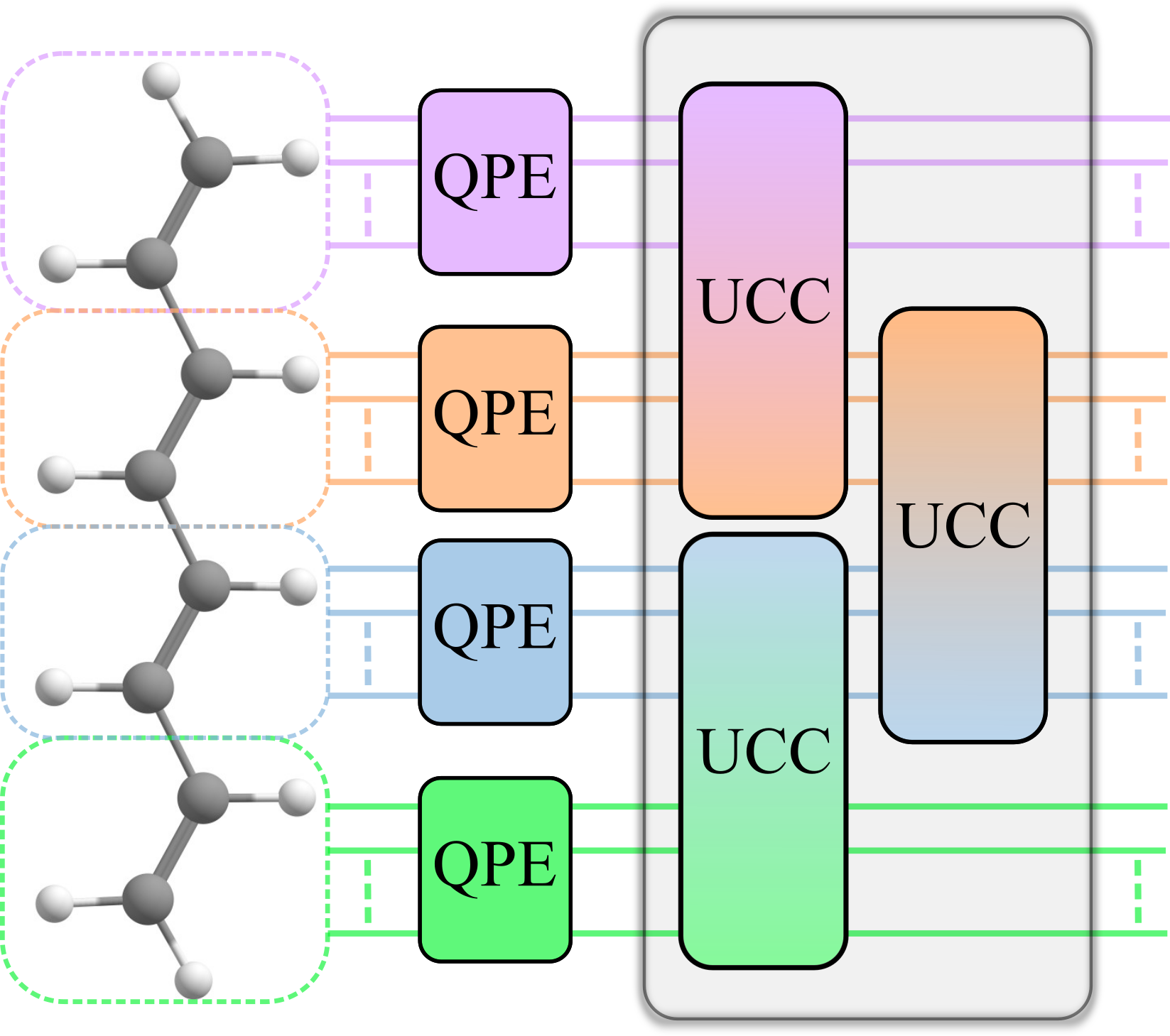}
    \caption{Diagram of example circuit using LAS-UCC. The system of interest is first separated into distinct fragments. QPE is used on each fragment to solve for the approximate unentangled ground state. Correlation between fragments is then added in, variationally, through a unitary coupled cluster ansatz.}
    \label{fig:circuit}
\end{figure}

A schematic representation of the described circuit is shown in Fig.~\ref{fig:circuit}.
This provides electron correlation between the fragments, in a way
that scales exponentially on classical computers, but only polynomially on
quantum computers. Moreover, this procedure provides a better 
estimate of the ground state energy than the product wave function or the UCCSD would
provide alone. Note that, unlike LASSCF, this method is not strictly variational (despite the use of VQE) because the initial product-state wave function, $\bigwedge_K \ket{\Psi_K}$, is not variationally reoptimized in the presence of the UCCSD correlators.
The QPE circuits could also be replaced with a local variational ansatz, leading to a fully variational algorithm, which we term LAS-VQE and describe in the Supplementary Information.

To understand the large improvement in computational complexity of our approach, 
we focus on a system of $n_f$
fragments, with the number of orbitals per fragment, $N_K$, constant as the number of fragments
grows. The total system size is defined by 
$N=N_K n_f$ orbitals. We also
assume that each fragment interacts with only the $m$ geometrically nearest fragments and
that $m$ does not grow with $n_f$. These are reasonable assumptions for many
interesting molecules and mirror the assumptions made in classical LASSCF.
Under these assumptions, the QPE solver for the unentangled fragments does not
grow with $N$, since $N_K$ is
assumed to be fixed while $n_f$ grows. 
The number of small QPE sections grows linearly with the number of fragments, of course.
Typically, the Jordan--Wigner transformation would introduce
an $O(N)$ term 
to enforce the anticommutation relations among the orbital creation
and annhilation operators.
However, in the case of 
linear chains, as we study here, ordering the orbitals such that all up and down occupied and virtual orbitals in a given fragment 
are close, the high-weight $Z$ part of the Jordan--Wigner transformation effectively cancels out, causing no scaling with total number of orbitals. See Supplementary Information for more details. Together, this
leads to an overall
$O( n_f N_K^4) \approx O(N)$ (linear) number of gates to solve for the
$n_f$ unentangled product wave functions.
The UCCSD correlator, which is then applied, has $O\big(m^4 N_K^4)$ terms in the cluster
operator for each correlator, because the UCCSD circuit  spans only $m$ fragments. 
Neither $m$ nor $N_K$ grows with the total size (number of spin orbitals) of the system, $N$. The number of $m$-local correlators grows as $O(n_f)$.
Again, by careful ordering of the orbitals, the
Jordan--Wigner transformation does not introduce any scaling overhead.
The complexity of the $m$-local UCCSD correlator is then 
$O(n_f m^4 N_K^4) \approx O(N)$ (linear). 
This creates an overall linear scaling in the number of gates
for linear chain geometries, with respect to only the total size of
the system, $N$, and is polynomially ($O(N^4)$) better than performing QPE alone, while
providing accuracy above VQE using the UCCSD ansatz and classical LASSCF. Many of the 
gates can be done in parallel,
such as the local QPE circuits and the 
different $m$-local UCCSD correlators, leading to an 
expected overall sub-linear depth. If the fragments are 
coupled in a geometry more complicated than a linear chain, the UCCSD
correlator will potentially incur the $O(N)$ Jordan-Wigner overhead,
leading to an overall $O(N^2)$ scaling for arbitrary geometries with
an expected $O(N)$ depth.


\subsection{Illustrative Molecular Systems}

\begin{figure}
    \centering
  \includegraphics[width=0.75\linewidth]{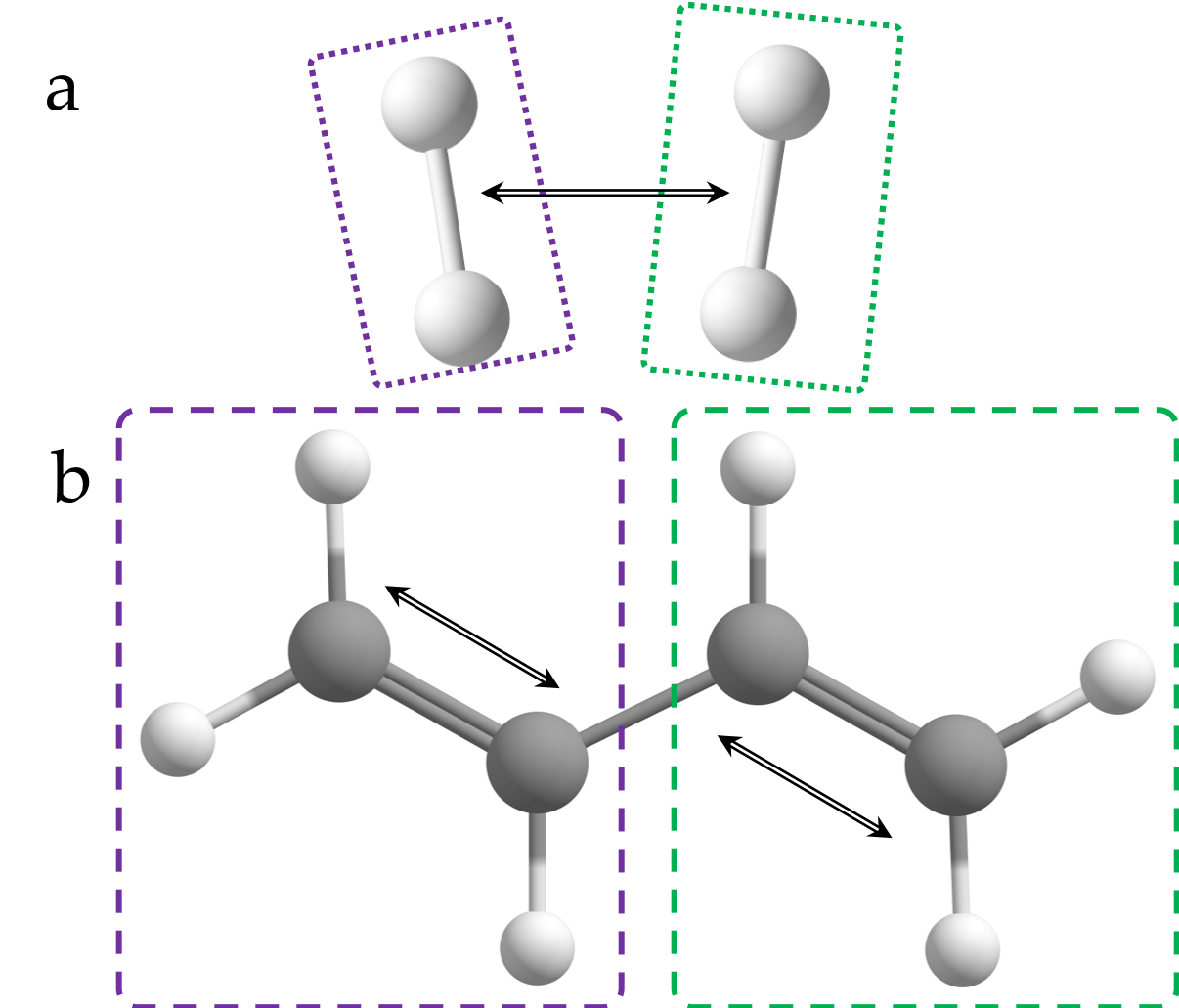}
    \caption{Two model systems used for testing. (a) The asymmetric hydrogen dimer,  $(\rm{H}_2)_2$. Each H$_2$
      molecule is a fragment described by a 2-electron, 2-spatial orbital or (2,2) active subspace in the dimer's LAS wave function. The potential
      energy surface is scanned along the distance between the two H$_2$ bond
      midpoints, indicated by the black double line.
      (b) The \emph{trans}-butadiene molecule at its CASSCF(8,8)/6-31G
      ground-state equilibrium geometry. Dashed boxes depict the two
      notional fragments containing the two (4,4) active subspaces in the LAS
      wave function. Black double lines indicate the internal coordinate along which
      the potential energy surface is scanned; the two terminal methylene units
      are simultaneously removed from the
      central acetylene unit.} 
    \label{fig:molecules}
\end{figure}

In the calculations discussed below, we consider two systems, depicted in
Fig.~\ref{fig:molecules}. The first, shown in Fig.~\ref{fig:molecules}(a),
is a simplistic model of weakly interacting fragments, consisting of two H$_2$
molecules at various distances between their two midpoints using a minimal STO-3G
atomic orbital (AO) basis set, and the two active subspaces in the LAS wave
function correspond to the active spaces of the two H$_2$ molecules. We use this small basis set because of the size limitations of today's quantum computers and simulations. The bond lengths and internal
angles of this system are set arbitrarily to remove point group symmetry so
that differences between various methods are not obscured by the simplicity of a
symmetrized electronic wave function. The interaction between the two fragments
in this model system are weak, and the LAS wave function is therefore expected to
provide an excellent model of the FCI wave function except when the distance
between the two molecules is very small. We additionally extend this system
up to 20 H$_2$ in a linear chain, where we  estimate only the 
total number of quantum resources necessary.

The second system, depicted in Fig.~\ref{fig:molecules}(b), is the
\emph{trans}-butadiene molecule. The
potential energy surface of this molecule is scanned along the internal
coordinate corresponding to the simultaneous stretching of both the C=C
double bonds, leading to the removal of two  methylene units from a
central C$_2$H$_2$ (distorted acetylene-like) unit. In the LAS wave function, the molecule is divided
into two fragments split across the central C--C bond, and each fragment
is described by a (4,4) active subspace. Several molecular orbitals are therefore
left inactive, described by an unfragmented single determinant. 
We employed the 6-31G AO basis set in this case.

The \emph{trans}-butadiene system is a chemical model of the case of two strongly interacting units in a system, where
the value of the stretching internal coordinate is a proxy for the strength of 
electron correlation. Near the equilibrium geometry, dividing the
active space into two fragments is chemically reasonable: each fragment encloses
one $\pi$-bond, and inasmuch as electron correlation affects the system at all,
it is a reasonable approximation to consider it only locally. However, as the
C=C double bonds are elongated, electrons from the two broken $\pi$ bonds
recouple across the central C$_2$H$_2$ unit, which spans the fissure between
the two LAS fragments. The LAS wave function cannot model a $\pi$ bond in this
position, and the LASSCF method breaks down.

\section{Results and Discussion \label{sec:results}}
\begin{figure}
    \centering
    \includegraphics[width=0.96\linewidth]{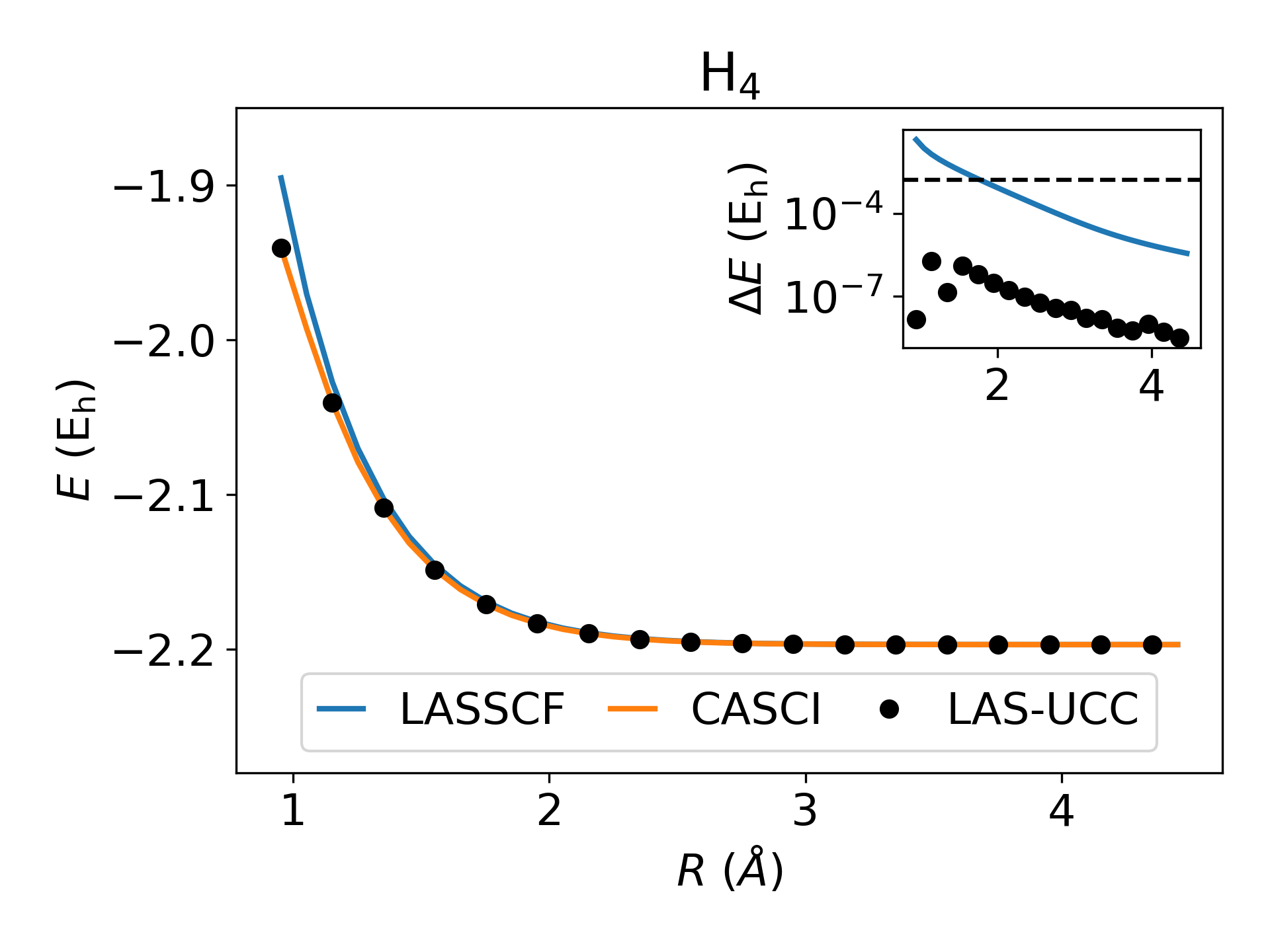}

    \caption{Energies for (H$_2$)$_2$ calculated by CASCI, LASSCF, and LAS-UCC. The inset shows the error, with respect to CASCI, of LASSCF and LAS-UCC. The black dashed line represents chemical accuracy.
    LAS-UCC is able to obtain chemical accuracy, with respect to CASCI, at all distances. LASSCF cannot obtain chemical accuracy at sufficiently short distances. }
    \label{fig:h4}
\end{figure}

\begin{figure}
    \centering
    \includegraphics[width=0.96\linewidth]{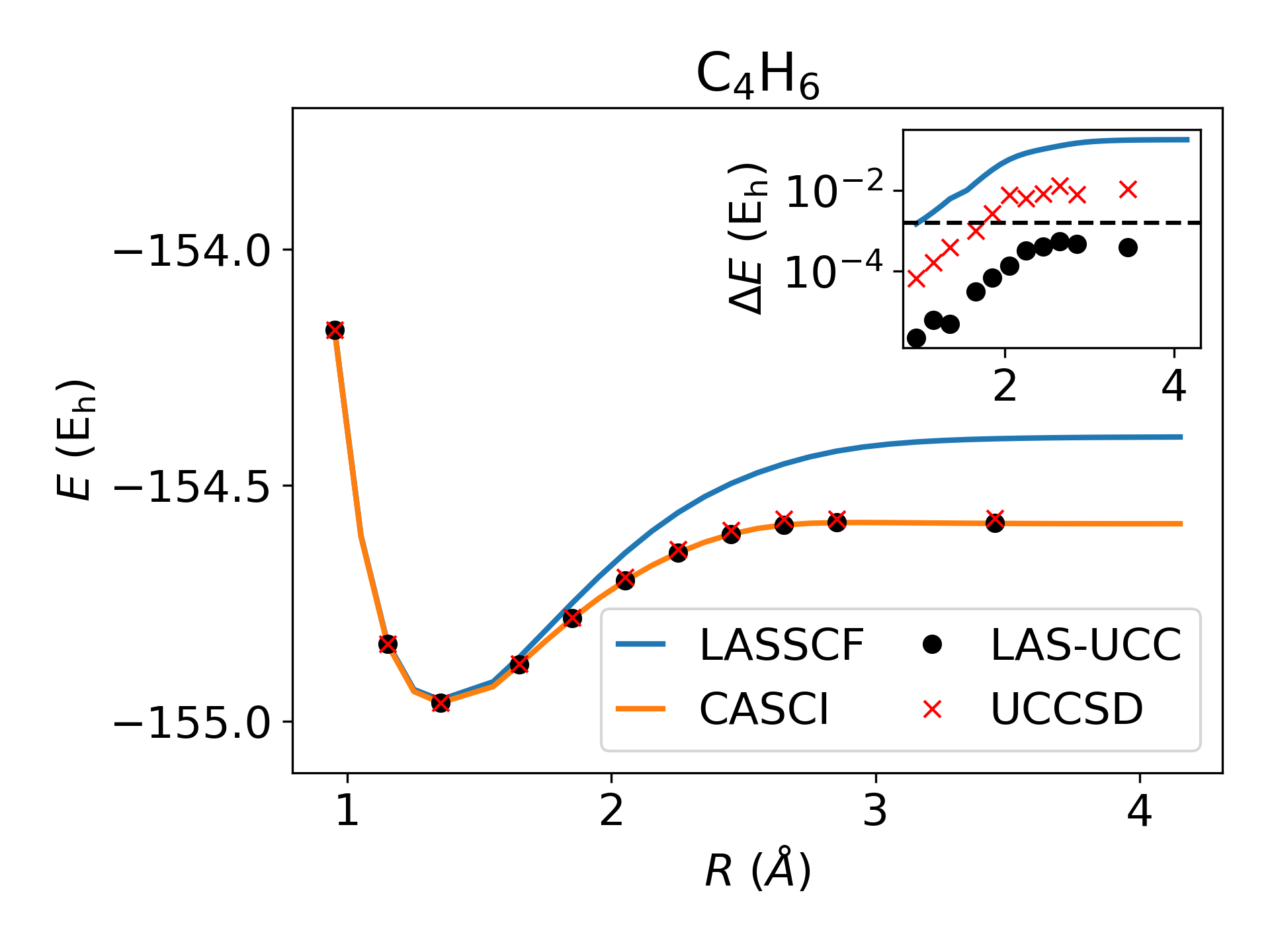}

    \caption{Energies for C$_4$H$_6$ calculated by CASCI, LASSCF, and LAS-UCC. The inset shows the error, with respect to CASCI, of LASSCF and LAS-UCC. The black dashed line represents chemical accuracy. LAS-UCC obtains chemical accuracy across the potential energy surface, whereas LASSCF, which cannot accurately represent the correlation between the fragments, fails to obtain chemical accuracy for most points.}
    \label{fig:c4h6}
\end{figure}

\subsection{LAS-UCC}
We demonstrate the efficacy of our framework by simulating the two benchmark
molecules, (H$_2$)$_2$ and \emph{trans}-butadiene, described above. 
We compare three methods: LASSCF, CAS configuration interaction in the basis of LASSCF orbitals (CASCI), 
and our new algorithm, LAS-UCC. LASSCF
represents the best unentangled set of wave functions and is equivalent to the
solution after the QPE circuits but before the use of the UCCSD ansatz. Note that
CASCI is slightly different from CASSCF since the orbitals are not variationally
reoptimized. CASCI solves for the FCI wave function within the active space; in
this case, it is equivalent to using QPE across the whole molecule
and represents the reference result in these studies.

Figure~\ref{fig:h4} shows the results of applying the methods to the hydrogen dimer as the two H$_2$ molecules are
pulled apart. We see that LASSCF, CASCI, and
LAS-UCC agree except for very small distances where LASSCF no longer provides accurate
energies. 

Figure~\ref{fig:c4h6} shows the results for \emph{trans}-butadiene,
a model of strongly correlated fragments. Here, as the terminal methylene units
are removed, the interfragment correlation grows as a double bond is formed
between the fragments. The UCCSD ansatz can accurately represent this level of
entanglement, allowing LAS-UCC to achieve nearly CASCI accuracy, whereas LASSCF
fails to account for this entanglement. \change{With a standard Hartree-Fock initial state, as is typically done in VQE, the UCCSD 
ansatz is unable to obtain chemical accuracy for the large distances. 
We also attempted to use the so-called `hardware-efficient' ansatz~\cite{kandala2017hardware},
but were unable to obtain results significantly better than Hartree-Fock using depths up to 10 (which corresponds to a similar 
number of parameters as the UCCSD ansatz) at equilibrium.}

\subsection{Resource Estimates}

To demonstrate the scaling advantage of our method, we perform resource estimation for the number
of logical quantum gates necessary for several different quantum algorithms: the QPE algorithm over the 
full unfragmented molecule; the UCCSD ansatz over the full unfragmented molecule; and the two 
steps of our proposed LAS-UCC method, the fragmented QPE 
and the 2-local UCCSD (which corresponds to
the circuit depicted in Fig.~\ref{fig:circuit}). We estimate the number of resources
needed for the QPE algorithm if only a single Trotter time step were needed; $O(1000)$ 
time steps will be needed for typical systems to get to chemical accuracy~\cite{alan_2020_review,
garnet_2020_review}. 
Note that these estimates represent only the number of two-qubit CNOT gates, which we use as a 
primary gauge of the number of total resources. 
Single-qubit gates are also necessary; the
estimates for these resources can be found in the Supplementary Information and scale similarly to the number of CNOT gates.
\change{We also note here that we are only comparing the scaling number of gates; 
QPE, with a sufficiently good initial state and enough Trotter states, will of course be the most accurate
of all compared algorithms.}

\begin{figure}
    \centering
    \includegraphics[width=0.95\linewidth]{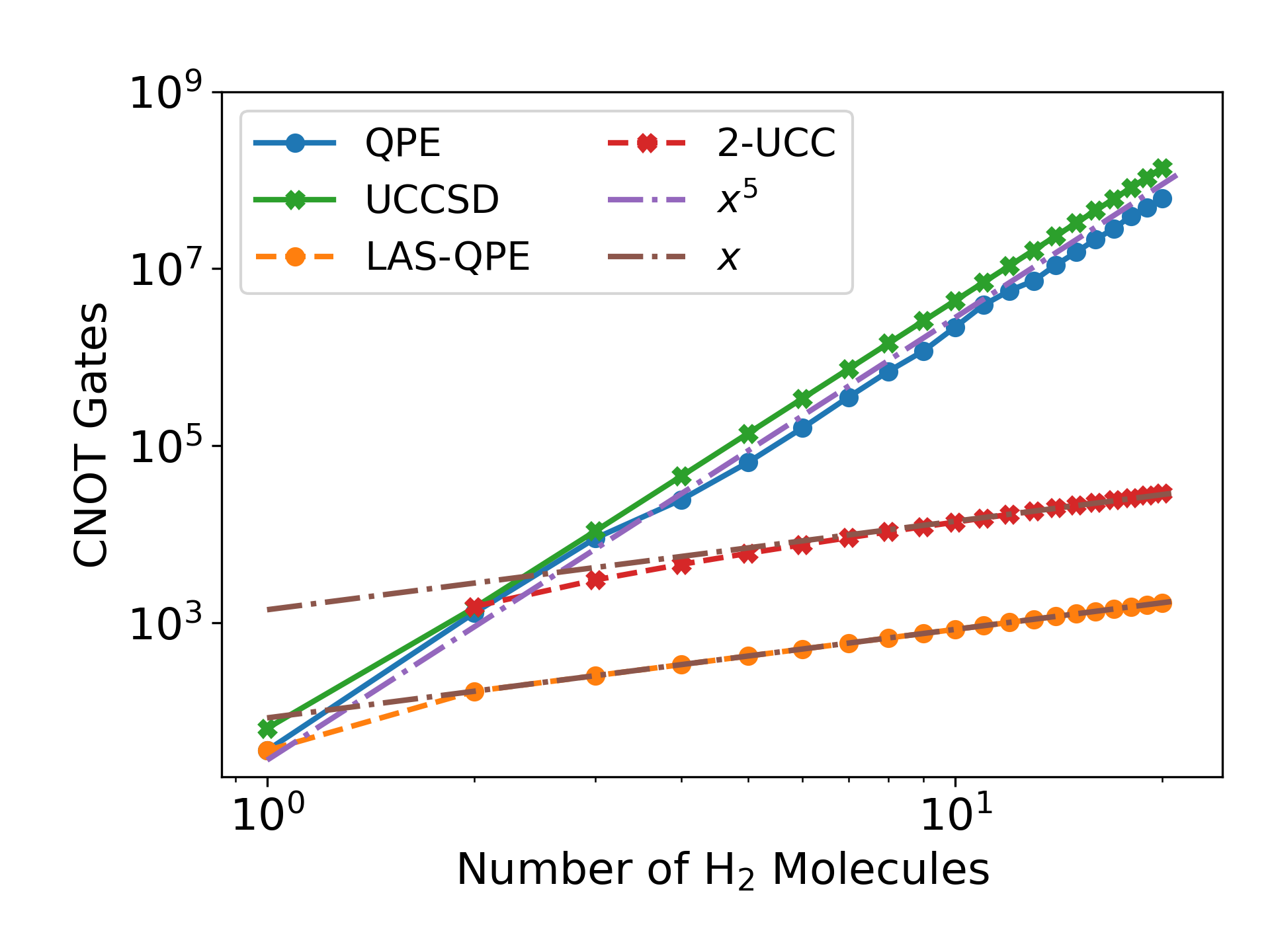}
    \caption{Estimated two-qubit gate counts using various algorithms. The QPE estimates assume only a single Trotter step; \change{$O(1000)$ will need to be taken to obtain chemical accuracy}.
    Polynomials of various orders have been plotted to demonstrate the scaling. Our algorithm, LAS-UCC, requires both the LAS-QPE and 2-UCC circuits and thus has an overall $O(N)$ scaling, compared with the $O(N^5)$ scaling of UCC and QPE.}
    \label{fig:gate_counts}
\end{figure}

We use a model system of 
an increasing number of H$_2$ molecules and look at how the number of CNOT gates increases
as the number of molecules increases, as shown in Fig.~\ref{fig:gate_counts}. 
As the number
of H$_2$ molecules increases, the number of gates needed for all methods also increases. As 
predicted in the complexity analysis of QPE [see Methods section], the
total number of gates for a single Trotter step in the QPE algorithm grows as $O(N^5)$.
Similarly, the number of gates needed for a global UCCSD ansatz also grows as $O(N^5)$, 
as expected~\cite{mcclean2017hybrid}. This result is compared with the much smaller number of gates necessary
to implement the two steps of our LAS-UCC algorithm. As expected, both the QPE and UCCSD 
parts of LAS-UCC provide dramatic scaling advantages, with the 2-local UCCSD ansatz and the QPE of the reduced Hamiltonian both scaling as only $O(N)$. 
We note that, in addition to evaluating the quantum circuits here, an
additional optimization loop is needed when using the UCCSD ansatz, whether it is
global or 2-local. Using a 2-local UCCSD ansatz also greatly reduces the number
of parameters that need to be optimized compared with a global UCCSD ansatz. 

\subsection{Discussion}
Here we compare LAS-UCC with the two quantum algorithms that it 
is composed of: QPE and variational UCCSD.
Compared with global QPE, LAS-UCC reduces the total quantum resource cost by 
approximating the system with noninteracting fragments and adding in some interaction between fragments (those described by a UCCSD ansatz spanning
the fragments). This in general reduces the accuracy; but as shown in the preceding sections, 
LAS-UCC provides accuracy comparable to CASCI (and therefore global QPE) for the 
systems considered here. The \emph{trans}-butadiene molecule is a model for larger, more complicated
systems of strongly interacting units. Many single molecular magnets have such pockets of strong correlation localized on the metal centers, which moderately interact with each other~\cite{doi:10.1021/ja0316824,baniodeh2018high}. With LAS-UCC we  not only can obtain the wave function efficiently but also can selectively couple the fragments with the UCC correlator, offering further insight into the nature of these interactions. Affordable and accurate modeling of phenomena such as singlet fission \cite{smith2013recent,casanova2018theoretical} in molecular crystals of conjugated organic compounds can be performed  with LAS-UCC, as fault-tolerant quantum computers become available. 
This approach will also be used to study chemical processes involving interfragment bond formation and breaking while still treating all points on a potential energy surface at comparable footing.


Compared with standard UCCSD, LAS-UCC can be seen as augmenting UCCSD with a 
multireference initial state. Instead of using single-determinant Hartree--Fock,
as is standard in VQE demonstrations of UCCSD~\cite{doi:10.1063/1.5133059,doi:10.1126/science.abb9811,mccaskey2019quantum,PhysRevX.6.031007,kandala2017hardware}, LAS-UCC uses the unentangled
product state of the ground state wave functions of each fragment (which is 
also the LASSCF wavefunction). This provides additional accuracy, 
above standard single-reference UCCSD, at a negligible increase in cost. 
When using a global UCCSD
ansatz, the increase in the number of gates is negligible, even when taking
into account the $O(1000)$ time steps that would be needed to implement the 
QPE step. Using the $m$-local ansatz provides further reduction.
\change{There have been other proposals for preparing interesting,
multireference initial state in context of efficiently finding
states with large overlap with the true ground state~\cite{tubman2018postponing,sugisaki2018quantum}. These 
algorithms could be used in-place of the QPE part of LAS-UCC to provide
the initial state and potentially
adapted to give similar LASSCF-like states with similar overhead reductions as shown for QPE.}

Moreover,  recent advances in VQE algorithms have developed various ways to reduce the cost associated with the UCC correlator~\cite{grimsley2019adaptive,alan_2020_review,cao2019quantum,cerezo2021variational,bharti2021noisy,fedorov2021unitary}.
As presented in the Theory section, LAS-UCC can also be seen as a post-LASSCF method that recouples select fragments at a level of theory beyond the mean field. 
The addition of the doubles or higher terms in the cluster operator provides a way to systematically improve the accuracy beyond the LASSCF reference. 
\change{On classical computers, such an approach requires truncating~\cite{Hoffmann1988} or approximating~\cite{Neuscamman2010} the non-terminating BCH expansion in a more or less arbitrary way.}

Not every system will be accurately described by LAS-UCC, of course, but one can systematically increase the accuracy in several ways, while 
increasing the total resource cost. Increasing the 
size of each fragment (which in turn decreases the number of fragments)
gradually increases the accuracy, until the limit of a single fragment,
where the UCCSD ansatz becomes redundant and the algorithm
becomes simply global QPE. On the UCC side, the order of the ansatz can be 
increased. Triples, quadruples, and so on can be included at increasing cost. If
using an $m$-local ansatz, the scaling is unaffected, but the total
number of gates increases. The locality of the ansatz, $m$, can also be increased,
providing explicit correlation between more geometrically distant fragments. 

\section{Conclusions \label{sec:conclusions}}

We introduced LAS-UCC, a quantum algorithm that combines a fragmentation of the wave function of a chemical system with QPE and variational UCCSD 
to compute the ground state
energy of such a system. LAS-UCC  can describe compounds containing strongly interacting
fragments, and it provides a polynomial scaling advantage in
the number of quantum gates compared with
other quantum algorithms such as QPE and UCCSD.  Since the fragments' reduced Hamiltonians have fewer terms and by ensuring the locality of the Jordan-Wigner transform, the overall gate count will be $O(N)$
with respect to the total size of
the system $N$ for  for linear geometries and $O(N^2)$ more generally,
compared with $O(N^5)$ requirements for QPE. We also demonstrated the
accuracy of LAS-UCC on (H$_2$)$_2$ and \emph{trans}-butadiene molecules and performed resource
estimations of larger systems to \change{provide evidence for potential scaling advantages.}

As larger fault-tolerant quantum computers are developed, we expect
that our algorithm will be able to provide accurate calculations of 
large and useful chemical systems, such as molecular magnets and qubits, photovoltaic materials, and large biomolecules that are out of reach of classical computing algorithms but 
for which QPE would be too expensive. 

\section{Methods}
\subsection{Quantum Algorithms}
Here we describe two quantum algorithms that  serve as the primary
components for our fragment-based quantum algorithm.

\subsubsection{Quantum Phase Estimation}
The quantum phase estimation  algorithm solves for the eigenvalue, $\lambda_k$, for
an eigenvector $|v_k\rangle$ of some unitary matrix, $U$. In addition to its
use in quantum chemistry, it forms the basis
for many important quantum algorithms, such as Shor's prime number factoring
algorithm~\cite{shor1999polynomial} and the Hassidim--Harrow--Lloyd algorithm for inverting
matrices~\cite{PhysRevLett.103.150502}. For quantum chemistry problems, the unitary matrix $U$ is
generated by the Hamiltonian, $H$ (eq.~\eqref{qchem_ham}), over time steps $\tau$:
\begin{equation}\label{qpe_phase}
  U |v_k\rangle = e^{-i\hat{H}\tau} |v_k\rangle = e^{i 2\pi \phi} |v_k \rangle,
\end{equation}
and the desired energy is mapped to the phase acquired,~$E=-2\pi \phi/\tau$, 
where units have been chosen such that $\hbar=1$. By combining real-time evolution
of the Hamiltonian, $\hat{H}$, with 
application of the quantum Fourier transform
(QFT)~\cite{365700}, the
value of the energy can be obtained in polynomial time using a quantum computer.

The computational complexity of the QPE is directly related to the complexity of
implementing the unitary propagator $U=e^{-i \hat{H} \tau}$.  Many strategies for implementing $U$ exist,
including Trotterization~\cite{PhysRevA.64.022319,PhysRevA.91.022311}, Taylorization~\cite{PhysRevLett.114.090502}, and
qubitization~\cite{Low2019hamiltonian}. The Hamiltonian, Eq.~\eqref{qchem_ham},
has $O(N^4)$ terms, where $N$ is the number of spin orbitals. Each term in the
Hamiltonian can be transformed into a Pauli string (that is,
a product of Pauli operators $X$, $Y$, $Z$, or $I$) via one of the many
fermion-to-spin transformations, such as the Jordan--Wigner~\cite{jordan1993paulische},
parity~\cite{bravyi2017tapering}, and Bravyi--Kitaev~\cite{BRAVYI2002210} transformations. In this
work we  focus on QPE using Trotterization with 
the Jordan--Wigner transformation since they serve as standard
reference points for the other variations. The complexity
of QPE for the Hamiltonian, Eq.~\eqref{qchem_ham},
using Trotterization with the Jordan--Wigner transformation is $O(N^5)$: 
$N^4$ arising from the number of terms in the Hamiltonian and
an additional $N$ from the Jordan--Wigner transform.
Although QPE can obtain estimates of the ground state energy with
only a polynomial number of quantum gates, the overheads are still too large for
near-term quantum computers. The success of the QPE algorithm directly depends on the
overlap of the initial state (which is often taken to be the 
Hartree-Fock state) and the true ground state. Realistic estimates, taking into account overheads
such as quantum error correction, put the needed number of qubits to perform QPE
on interesting molecules in the millions~\cite{elfving2020will,liu2021prospects,kim2021fault}.

QPE is analogous to a Fourier analysis of a correlation               
function; and, for a given energy accuracy, $\epsilon$, it requires propagation
efforts (maximum times) on the order of $O(1/\epsilon)$~\cite{alan_2020_review,
garnet_2020_review}. Since the circuit depth for evaluating the propagator
for individual fragments will naturally be lower than for the full system,
the QPEs involved in our LAS approach will be significantly cheaper
than full QPE.  

\subsubsection{Variational Quantum Eigensolver}
The variational quantum eigensolver  is a hybrid quantum-classical
algorithm that relies on the variational principle to find an estimate of the ground state
energy of a given molecule. A circuit with variable parameters, $\theta$,
serves as an ansatz, whose energy is evaluated on a quantum computer and whose
parameters are iteratively optimized by a classical computer. For a circuit ansatz
$|\psi(\theta)\rangle$, VQE estimates the energy as
\begin{equation}\label{eq:vqe2}
  E = \min_{\theta} \braket{ \psi(\theta) | \hat{H} |\psi(\theta)  }.
\end{equation}
The Hamiltonian, $\hat{H}$, is transformed into a sum of Pauli strings via a
fermion-to-spin transformation, and
the expectation value of each term is measured from the quantum computer 
separately and summed on the classical computer. VQE has
much less stringent quantum resource requirements than QPE has, since it
offloads much of the work (such as optimization) to the classical computer. Hence, VQE has been used in proof-of-principle calculations for small
molecules~\cite{doi:10.1063/1.5133059,mccaskey2019quantum,PhysRevX.6.031007}.

The accuracy of VQE is determined by the quality of the ansatz,
$|\psi(\theta)\rangle$. The UCCSD
ansatz is an interesting choice as wave function for VQE since there is no known way to efficiently implement UCCSD on classical computers~\cite{BARTLETT1989133,taube2006new,kutzelnigg1991error}, but it can be implemented with $O(N^5)$
gates on quantum computers~\cite{PhysRevA.95.020501,doi:10.1063/1.5011033,mcclean2017hybrid}. The UCCSD ansatz is
\begin{equation}
  \ket{\psi_{{\rm UCCSD}}} = \hat{U}_{{\rm UCCSD}} \ket{{\rm HF}} = \exp{\hat{T}_{\rm UCCSD}} \ket{{\rm HF}},
\end{equation}
where $\hat{T}_{\rm UCCSD}$ is defined by truncating the more general
cluster operator of Eq.\ (\ref{eq:tcc}) at the second term. While the UCCSD ansatz can be implemented 
on NISQ devices for small molecules~\cite{doi:10.1126/science.abb9811,kandala2017hardware}, it is limited
in its accuracy because of only including up to doubles excitations.

\subsection{Computational Methods}
To calculate the accuracy of the proposed method for small molecules, 
we use the following strategy. We first use a classical LASSCF solver,
as implemented in the \textit{mrh} package~\cite{mrh_software}, to find the best product wave function.
This effectively provides an equivalent solution to that of the QPE step
of our proposed algorithm. We then represent this product wave function as a CI vector in the complete active Fock space and apply a UCCSD correlator, as well as its derivatives with respect to all amplitudes, to this reference CI vector. We employ the factorization reported by Chen et al.~\cite{Chen2020b} to avoid the BCH expansion and its inevitable approximate truncation. The resulting $\ket{{\rm QLAS}}$ CI vector and its derivatives ($\ket{\delta {\rm QLAS}}$) with respect to the unitary coupled cluster amplitudes are used to compute the energy, $\braket{{\rm QLAS}|\hat{H}|{\rm QLAS}}$, and its derivatives, $\braket{\delta{\rm QLAS}|\hat{H}|{\rm QLAS}}$. We then minimize the former using the latter and the Broyden–-Fletcher-–Goldfarb-–Shanno algorithm.
We find that this approach is more
efficient than directly simulating the quantum circuits. We note that this method
scales exponentially on classical computers.

To provide gate count estimates, we use the Q\# package~\cite{svore2018q}, generally
following the framework of Ref.~\onlinecite{low2019q}. The 
full and reduced Hamiltonians are produced by using the \textit{mrh} package~\cite{mrh_software}, and 
both Hamiltonians are then passed to the Q\# package to estimate the number of CNOT gates using
the QPE algorithm with a single Trotter time step for each. 
Additionally, we estimate the number of CNOT gates 
necessary to calculate various UCCSD ansatzes, including a global
UCCSD ansatz over the whole unfragmented molecule and multiple 2-local ansatzes
that span only two fragments. We count only the number of logical quantum 
gates needed. Real quantum computers will require additional overheads, 
owing to limited connectivity and the need to use expensive quantum error
correction protocols to deal with inevitable errors~\cite{liu2021prospects,kim2021fault}. Furthermore, 
we provide gate counts only; no attempt was made to count gate depth, which
is typically smaller, because many gates can be implemented in parallel.

\section{Contributions}
L.G., S.G. M.O. and M.R.H. designed the project. M.O. wrote the quantum algorithm and tested it. M.R.H. wrote the LASSCF classical code. R.P. tested the codes and performed some of the calculations. Y.A. helped with the theory and suggested testing calculations. M.O. and M.R.H. wrote the initial draft of the manuscript.
All authors contributed to the scientific discussions and manuscript revisions.

\begin{acknowledgments}
This research is based on work supported by Laboratory Directed Research and Development (LDRD) funding from Argonne National Laboratory, provided by the Director, Office of Science, of the U.S. DOE under Contract No. DE-AC02-06CH11357. 
This work was performed, in part, at the Center for Nanoscale Materials, a U.S. Department of Energy Office of Science User Facility, and supported by the U.S. Department of Energy, Office of Science, under Contract No. DE-AC02-06CH11357.
MRH and LG are partially supported by the U.S. 
Department of Energy (DOE), Office of Basic Energy 
Sciences, Division of Chemical Sciences, Geosciences, and 
Biosciences under grant no. USDOE/DE-SC002183.
This material is based upon work supported by the U.S. Department of Energy, Office of Science, National Quantum Information Science Research Centers.
We gratefully acknowledge the computing resources provided on Bebop, a high-performance computing cluster operated by the Laboratory Computing Resource Center at Argonne National Laboratory and University of Chicago Research Computing Center.

\end{acknowledgments}

\bibliography{library.bib}

\end{document}


\title{Supplementary Information for Localized Quantum Chemistry on Quantum Computers}
\date{\today}

\author{Matthew \surname{Otten}}
\email{mjotten@hrl.com}
\affiliation{HRL Laboratories, LLC, 3011 Malibu Canyon Road,
Malibu, CA 90265}
\author{Matthew R.\ \surname{Hermes}}
\affiliation{Department of Chemistry, Pritzker School of Molecular Engineering, James Franck Institute, Chicago Center for Theoretical Chemistry, University of Chicago, Chicago, IL 60637, USA.}
\author{Riddhish \surname{Pandharkar}}
\affiliation{Department of Chemistry, Pritzker School of Molecular Engineering, James Franck Institute, Chicago Center for Theoretical Chemistry, University of Chicago, Chicago, IL 60637, USA.}
\author{Yuri Alexeev}
\affiliation{Computational Science Division, Argonne National Laboratory, Lemont, IL 60439, USA}
\author{Stephen K.\ \surname{Gray}}
\affiliation{Center for Nanoscale Materials,
Argonne National Laboratory, Lemont, IL 60439, USA}
\author{Laura \surname{Gagliardi}}
\affiliation{Department of Chemistry, Pritzker School of Molecular Engineering, James Franck Institute, Chicago Center for Theoretical Chemistry, University of Chicago, Chicago, IL 60637; Argonne National Laboratory, Lemont, IL 60439, USA.}

\maketitle

\section{LAS-VQE}
Here we describe a more approximate approach than LAS-UCC wherein the QPE circuits for the fragments are replaced by UCC ansatzes, leading to a fully variational method.
The chemical knowledge that guides us in defining subspaces of the LAS wave function can also be used to reduce the size of the unitary operator for the UCC ansatzes. It suggests that operators corresponding to the higher excitation from one fragment to another do not affect the wave function significantly. Thus, we introduce a modified
ansatz for truncated UCC circuits where we  consider the “locality” of the excitation. An excitation involving only orbitals localized on a particular fragment---that is defined by the user---is classified as a “local excitation.” In
this modified UCC ansatz the user  not only can truncate the UCC excitation to a
certain maximum number (doubles, triples, etc.) but also can impose a constraint of
locality on the higher excitations. 
We develop here the theory and the algorithm of this method and name it LAS-VQE. In LAS-VQE, all the singles
excitations are included while only the local ones are included for the higher
excitations. This corresponds to having a mean-field interfragment interaction
and allowing all possible orbital rotations. 
This does not reduce the number of qubits required to represent the system, but
it does lower the complexity of the circuit as discussed below. 
In principle, one could use the locality argument to  include only  local
excitations, allowing us to reduce the number of qubits required. This separate-fragments VQE approach, however, fails to account for any interaction between the fragments. 
We choose to always include the singles excitations (orbital rotations) across
the entire system to account for some interfragment correlation and to make the
method less sensitive to the localization schemes used.

We also choose to perform in the singles excitation among spin-restricted
orbitals. We achieve this by using the same parameter to control the alpha and
beta excitations corresponding to a given pair of spatial orbitals. This is
necessary to avoid artificial lowering of energies due to spontaneous symmetry
breaking (the spin unrestricted solution). An example of the effect of this is
seen for the hydrogen dimer shown in Figure~2 of the main text. The nonspin adapted
UCCS energy is about 0.2 mHartree lower than the Hartree--Fock and has an spin
contamination of 0.09, showing the symmetry breaking. Working with
spin-restricted orbitals not only allows us to prepare symmetry preserving
states for the subsequent excitations, but also lowers the number of parameters
at the singles level by a factor of 2. This becomes a significant advantage
for the classical optimizer in VQE as the size of the system gets larger. 
\begin{figure}
    \centering
    \includegraphics[width=0.95\linewidth]{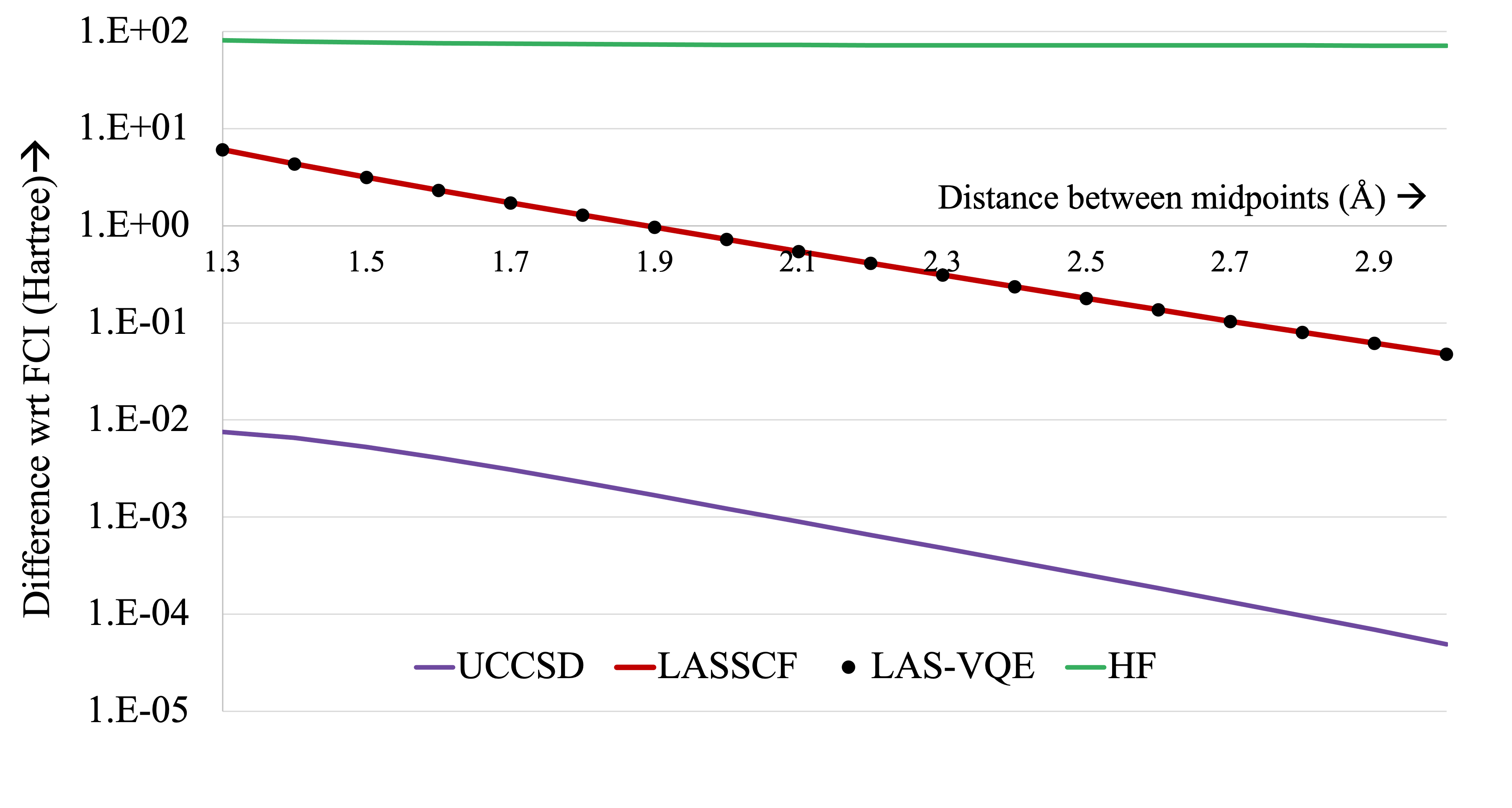}
    \caption{Energy difference (in Hartree) between FCI and the approximate methods as a function of the distance between the midpoints of the two H$_2$ molecules}
    \label{fig:las_vqe_log_scale}
\end{figure}
\begin{figure}
    \centering
    \includegraphics[width=0.95\linewidth]{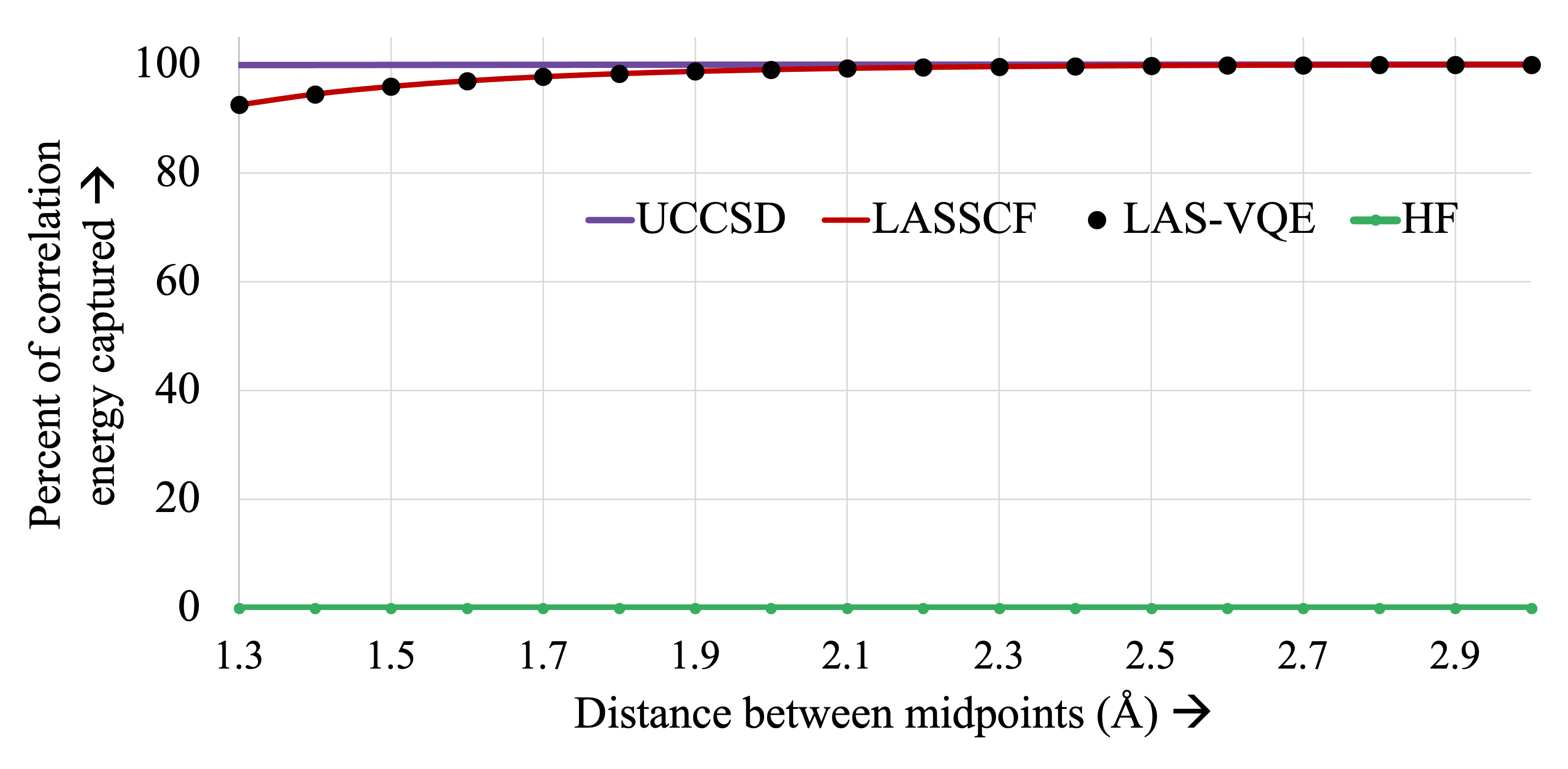}
    \caption{Percentage of correlation energy accounted for by the various methods for different intermolecular distances}
    \label{fig:percent_corr}
\end{figure}
We study the ground state electronic structure of the hydrogen dimer in various
conformations with the STO-3G basis set. We use CASSCF with a (4,4) active
space  and LASSCF with a ((2,2),(2,2)) active space, UCCSD and LAS-VQE. Each H$_2$
molecule is considered as one fragment on which the molecular orbitals are
localized. As one would expect, the performance of LASSCF with respect to CASSCF
(which is also FCI in this case) worsens with decreasing distance between the
two H$_2$ molecules. Figure~\ref{fig:las_vqe_log_scale} shows the energy difference between the various
methods and the FCI/CAS(4,4) absolute energies. UCCSD in this case is not
exactly FCI but does give  accurate results. LAS-VQE is close to
(but slightly lower than) the LASSCF energy. The LAS-VQE energy is not
variationally bound by the LASSCF that it is based on. 
In contrast to the LAS-UCC method discussed in this paper, this method uses a
single determinant 
Hartree--Fock reference state. The UCCSD energy is a variational lower bound to
the LAS-VQE method. There is, however, a significant advantage when it comes to
the computational cost:  gate count and the depth of the circuit. The number of
doubles operators scales linearly with the number of fragments. 
The Table~\ref{tab:lasvqe_cost} shows the number of gates and parameters required for the
various calculations for a system with increasing number of H$_2$ molecules with
the same fragmentation scheme. LAS-VQE in considerably
lower in both circuit depth and number of parameters,
compared with full UCCSD.

\begin{table}[ht]
	\centering
	\caption{Circuit depth and number of parameters for the various methods for an increasing number of H$_2$ fragments }
	\begin{tabular}{|c|c|c|c|c|c|c|}
	\hline
 &\multicolumn{3}{|c|}{Circuit Depth} & \multicolumn{3}{|c|}{Parameters} \\ \cline{2-7}
 & UCCS & LAS-VQE & UCCSD & UCCS & LAS-VQE & UCCSD \\ \hline
1 $\times$ H$_2$ & 14 & 97 & 97 & 1  & 2 & 2 \\ \hline
2 $\times$ H$_2$ & 58 & 141 & 1791 & 4  & 6 & 22 \\ \hline
3 $\times$ H$_2$ & 147 & 230 & 11206 & 9  & 12 & 108 \\ \hline
4 $\times$ H$_2$ & 296 & 379  & 42789 & 16  & 20 & 344 \\ \hline

	\end{tabular}
	\label{tab:lasvqe_cost}
\end{table}
\change{

\section{Justification of Eq. (9) of the main text}
Substituting Eqs. (4) and (6) into the Eq. (3) of the main text and differentiating with respect to generator amplitude $x_{\vec{k}}$ yields
\begin{equation}
    \pdv{E_{\textrm{LAS}}}{x_{\vec{k}}} = \bra{\Phi}\bigwedge_{L\neq K}\bra{\Psi_L}\wedge\braket{\vec{k}|\hat{H}-E_{\textrm{LAS}}|\Psi_K}\bigwedge_{M\neq K}\ket{\Psi_M}\wedge\ket{\Phi},
\end{equation}
the vanishing of which corresponds to the minimization of the LAS energy given by Eq.\ (3). Within the Hilbert space of the $K$th fragment, the energy minimization conditions for the $x_{\vec{k}}$ amplitudes corresponds to an eigenproblem,
\begin{equation}
    \hat{H}_K\ket{\Psi_K}=E_{\textrm{LAS}}\ket{\Psi_K},
\end{equation}
where
\begin{eqnarray}
    \hat{H}_K &\equiv& \bra{\Phi}\bigwedge_{L\neq K}\bra{\Psi_L}\hat{H}\bigwedge_{M\neq K}\ket{\Psi_M}\wedge\ket{\Phi}
    \nonumber\\ &=& \tilde{h}^{k_1}_{k_2} \crop{k_1}\anop{k_2} + \frac{1}{4} h^{k_1k_3}_{k_2k_4} \crop{k_1}\crop{k_3}\anop{k_4}\anop{k_2},
\end{eqnarray}
where $\tilde{h}^{k_1}_{k_2}$ is given by Eq.\ (10) of the main text. This effective Hamiltonian describes the $K$th fragment without interacting with any other fragment; $\hat{H}_K$ for various $K$ mutually commute.

For a system composed of noninteracting fragments $A$ and $B$ with Hamiltonians $\hat{H}_A$ and $\hat{H}_B$, energies $E_A$ and $E_B$, and wave functions $\ket{\Psi_A}$ and $\ket{\Psi_B}$, it is generally true that
\begin{equation}
    \left(\hat{H}_A+\hat{H}_B\right)\ket{\Psi_A}\wedge\ket{\Psi_B} = \left(E_A+E_B\right)\ket{\Psi_A}\wedge\ket{\Psi_B}.
\end{equation}
Noting that $\hat{H}_{\textrm{eff}}$ given by Eq.\ (9) of the main text is obviously
\begin{equation}
    \hat{H}_{\textrm{eff}}=\sum_K \hat{H}_K,
\end{equation}
we conclude that 
\begin{equation}
    \hat{H}_{\textrm{eff}}\bigwedge_K\ket{\Psi_K} = n_f\times E_{\textrm{LAS}}\bigwedge_K\ket{\Psi_K}.
\end{equation}
In other words, $\hat{H}_{\textrm{eff}}$ models interacting fragments as non-interacting subsystems described by unrelated, mutually commuting effective Hamiltonian terms. Thus, the QPE algorithm applied to $\hat{H}_{\textrm{eff}}$ generates $\ket{\textrm{QLAS}}\equiv \bigwedge_K\ket{\Psi_K}$ on the quantum circuit.
}
\section{Orbital Ordering and the Jordan-Wigner Transformation}
The Jordan-Wigner transformation~\cite{jordan1993paulische} is one of the many ways of transforming fermion operators into spin operators, and is a necessary 
step for performing quantum chemistry calculations on quantum computers.
Here, we very briefly discuss the overheads associated with the Jordan-Wigner
transformation and how it can be mitigated in certain fragmented geometries.

The Jordan-Wigner transformation transforms a fermion creation operator for
orbital $j$, $a_j^\dagger$, out of a total of $N$ spin orbitals as follows
\begin{equation}
    \tilde{a}_j^\dagger = Z^{\otimes j - 1} \otimes \frac{X-iY}{2} \otimes I^{\otimes N-1}, 
\end{equation}
where $\tilde{a}_j^\dagger$ is the transformed operator; 
$X$, $Y$, and $Z$ are the Pauli operators; and the notation $Z^{\otimes N}$
denotes applying the tensor operation, $\otimes$, $N$ times for Z (i.e., $Z^{\otimes 4} = Z \otimes Z \otimes Z \otimes Z$. The term $Z^{\otimes j - 1}$ ensures that
the transformed operators obey the correct fermionic anti-commutation relations.
In standard quantum algorithms for quantum chemistry, these $Z$ strings introduce 
very high-weight (that is, having many terms that are not $I$) terms into the Hamiltonian
and cluster operators, leading to an $O(N)$ overhead. To see this, 
we focus on just the one-body terms.
Under the Jordan-Wigner transformation, and assuming $j>k$,
\begin{equation}\label{eq:one-body}
    \tilde{a}_j^\dagger \tilde{a}_k = I^{\otimes k -1} \otimes \frac{Z(X-iY)}{2} \otimes Z^{\otimes j-k-1} \otimes\frac{X-iY}{2} \otimes I^{\otimes N-j}.
\end{equation}
The weight (number of non-identity terms) of this operator is $j-k+1$.
In the most general case, there will exist terms where spin orbital $1$ and $N$ will have a nonzero component, and the resulting weight is $N$, which introduces
the $O(N)$ overhead normally associated with the Jordan-Wigner transformation. 
There are many techniques for generally reducing this overhead~\cite{tranter2018comparison,PhysRevA.95.032332}. In our simple
chain of H$_2$ molecules, we can remove this scaling by choosing a particular
ordering of the orbitals. One typical ordering of the orbitals 
is the following: all occupied spin up orbitals, all virtual spin up orbitals,
all occupied spin down orbital, all virtual spin down orbitals. If this
ordering is followed when using the local actives spaces of each fragment, 
the $O(N)$ overhead is still applicable. Take, for example, just the one-body term
from the first occupied spin up orbital of the first fragment 
to the first virtual spin up orbital
for the first fragment. In this case, $j=1$ and $k=O(N)$, leading to the 
typical $O(N)$ Jordan-Wigner scaling. If, instead, the orbitals are ordered
with all occupied spin up orbitals of fragment one, followed by all virtual 
spin up orbitals of fragment one, followed the same for the second fragment, 
and so on, the overhead is removed. Now, $j=1$ and $k=O(N_k)$, where $N_k$ 
is the number of spin orbitals in each fragment. Since $N_k$ is assumed to be
fixed as the number of fragments grows, the $O(N)$ scaling is reduced to a constant.
To remove the Jordan-Wigner overhead from mixed spin two-body terms, the ordering
can be further changed so that each spin up orbital is immediately followed
by its analogous spin down orbital. This shows that the Jordan-Wigner $O(N)$
overhead can be removed from all terms in the reduced Hamiltonian, removing the
$O(N)$ scaling from the QPE part of LAS-UCC. For UCCSD correlators, the Jordan-Wigner overhead can be removed for certain geometries, such as the linear H$_2$ chain
studied in the main text. More complex geometries, such as higher dimensional
lattices, will re-introduce scaling terms, as the orbitals cannot be simultaneously
ordered to give low-weight Pauli strings for terms which couple both in the
multiple dimensions, such as $x$ and $y$ in a 2D lattice.

\section{Single Qubit Rotation Gates}
\begin{figure}
    \centering
    \includegraphics[width=0.95\linewidth]{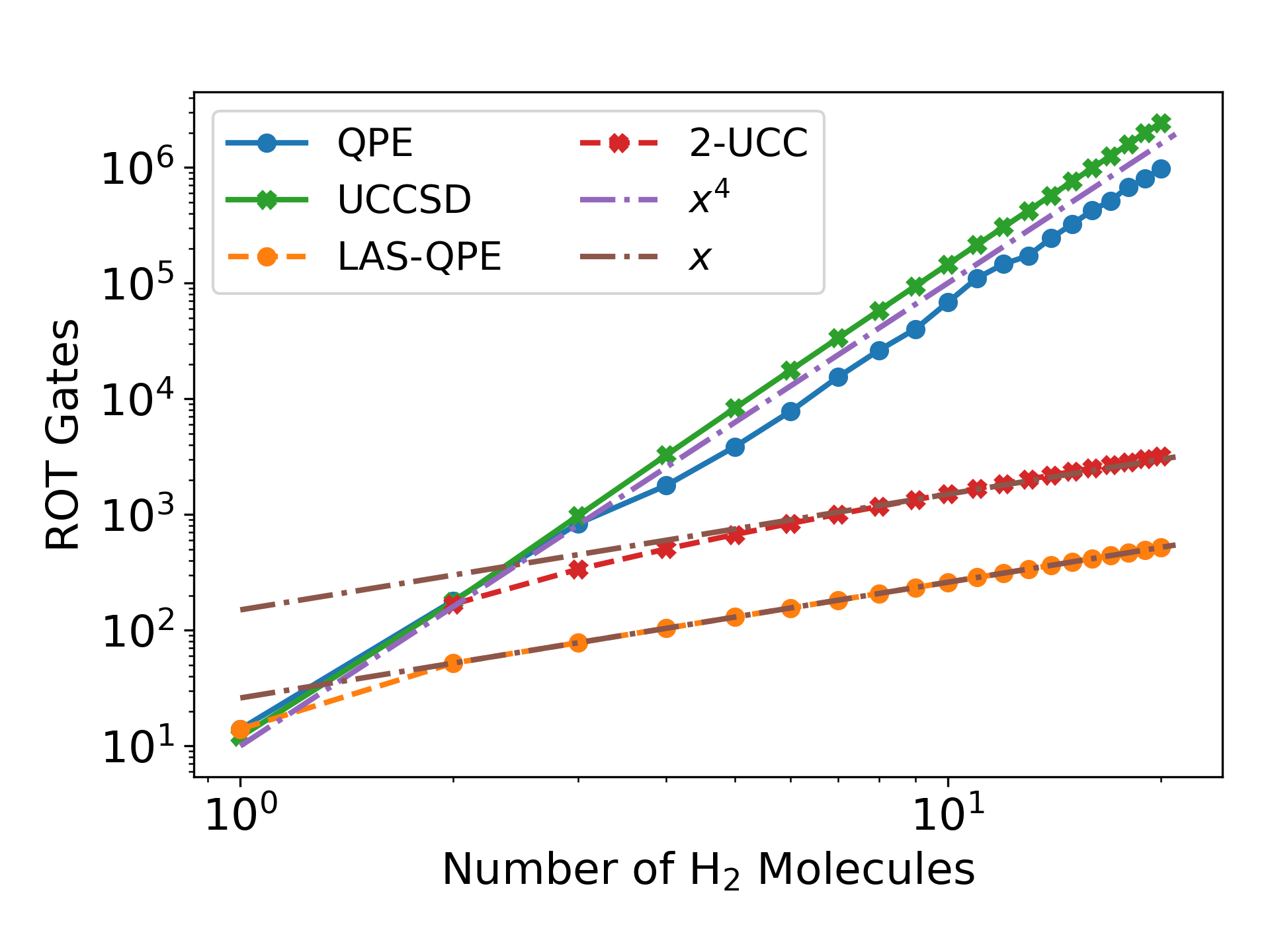}
    \caption{Estimated single qubit arbitrary rotation gate counts using various algorithms. The QPE estimates assume only a single Trotter step. Polynomials of various orders have been plotted to demonstrate the scaling. Our algorithm, LAS-UCC, requires both the LAS-QPE and 2-UCC circuits and thus has an overall $O(N)$ scaling, compared with the $O(N^4)$ scaling of UCC and QPE.}
    \label{fig:rot_gates}
\end{figure}

Figure~\ref{fig:rot_gates} shows the estimated number of arbitrary
single qubit rotation gates needed to implement full QPE across the 
whole molecule, the fragmented LAS-QPE, full UCCSD, and the 2-local
UCCSD correlator with increasing number of H$_2$ molecules. Just
like in the main text, LAS-UCC has polynomially fewer gates, 
compared with QPE or UCCSD, scaling only linearly for the
linear chain geometry. QPE and UCCSD only need $O(N^4)$ single
qubit rotations because the Jordan-Wigner overhead only shows up in the 
CNOT gates.

\change{
\section{Strong correlation of \emph{trans}-butadiene}
}
\begin{figure}
    \centering
    \includegraphics[width=1.0\columnwidth]{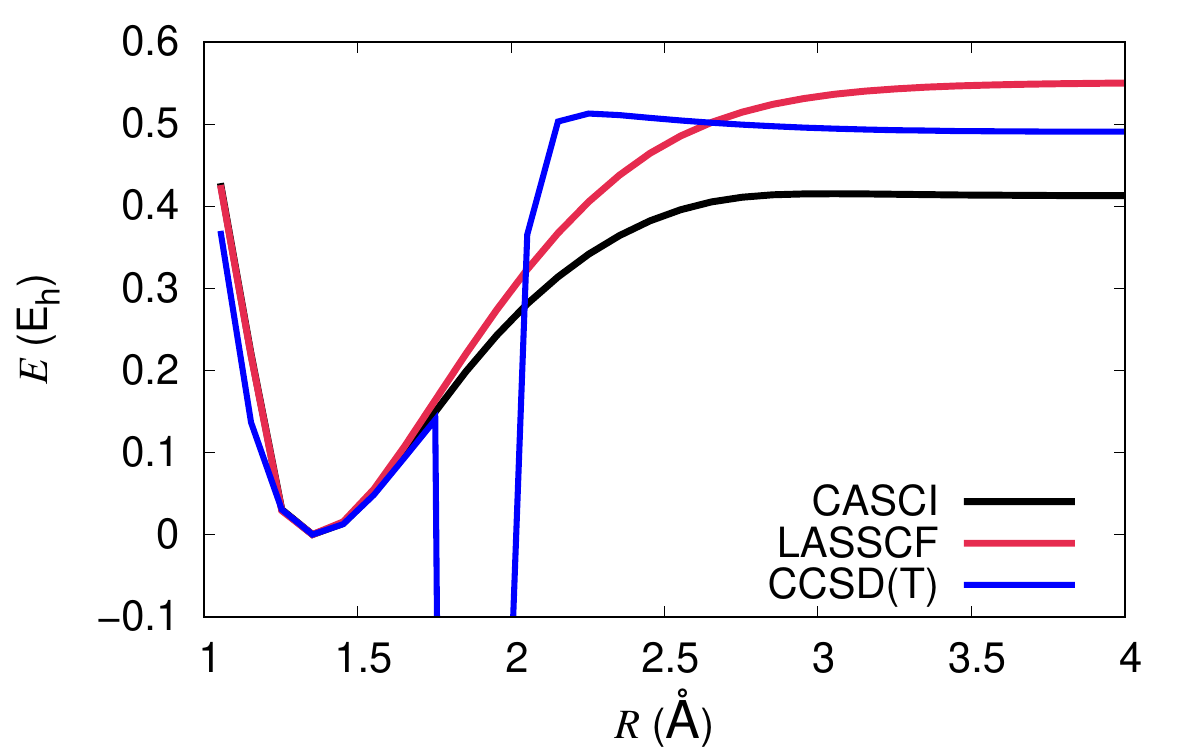}
    \caption{\change{Potential energy curve of the \textit{trans}-butadiene molecule along the symmetric double C=C bond dissociation coordinate, computed with CASCI, LASSCF, and CCSD(T). Total energies for each method are shifted to zero at the equilibrium geometry.}}
    \label{fig:c4h6_si}
\end{figure}
\change{
The symmetric double-bond dissociation curve of the \textit{trans}-butadiene molecule (cf.\ Figs.\ 2b and 4 of the main text) calculated using the CCSD(T) method is plotted in Fig.\ \ref{fig:c4h6_si} and compared to CASCI and LASSCF curves. In the dissociation limit, in which the molecule objectively can be modeled as three non-interacting diradical fragments, the CCSD(T) method recovers a greater amount of correlation energy than LASSCF, \textit{albeit} not as much as LAS-UCC or CASCI. Near the equilibrium geometry, the three methods are indistinguishable.

However, unlike LASSCF, CASCI, or LAS-UCC, all of which are multireference methods, the single-reference CCSD(T) method cannot smoothly connect the region of the equilibrium geometry to the bond dissociation limit. The discontinuous and unphysical behavior of CCSD(T) in the region of $R=1.8 \AA$ is consistent with the tendency of single-reference coupled cluster at low truncation orders to struggle to dissociate multiple or high-order bonds.\cite{Li2001} This \textit{trans}-butadiene potential energy curve is an example of a system in which both static and dynamical correlation must simultaneously be accounted for in order to obtain quantitatively accurate results.

\section{UCCSD operator ordering in LAS-UCC}

The following listings from the \textit{mrh} package describe the operator order for the first-order Trotterized UCCSD correlator used in our classical LAS-UCC calculations. Individual spatial orbitals are indexed first by fragment, and then in decreasing order of their natural orbital occupancy (before application of the UCCSD correlator). Listing \ref{lst:spatexc} contains the code for generating the combinations of spatial-orbital excitation patterns, as well as a docstring summarizing the effects of the code presented in Listings \ref{lst:spinexc} and \ref{lst:spincases} for the case of UCCSD. Listings \ref{lst:spinexc} and \ref{lst:spincases} contain the code for generating valid spin-orbital excitation patterns for any general set of spatial-orbital excitation patterns of any order in a UCC operator. Note that ``np'' is numpy and ``combinations\_with\_replacement'' is from the itertools module. This code is from the commit identified by the SHA-1 hash 8bb3c2d46ab5f43c8787a365b22713b5410855f2, which differs from the commit which performed the calculations reported in the main text (fe564cd1c54c404edf8bce1dd7118b2d53acdd7a) only in docstrings, test scripts, and line wrapping.
}
\begin{lstlisting}[caption={\change{Code for generating the spatial-orbital list of the UCCSD correlator in LAS-UCC within \textit{mrh}}}, language=Python, float=*, label=lst:spatexc]
def get_uccsd_op (norb, t1=None, t2=None):
    ''' Construct and optionally initialize semi-spin-adapted unitary CC
        correlator with singles and doubles spanning a single undifferentiated
        orbital range. Excitations from spatial orbital(s) i(, j) to spatial
        orbital(s) a(, b) are applied to the ket in the order

        U|ket> = u^n(n-1)_nn u^n(n-1)_n(n-1) u^n(n-2)_nn ... u^11_22 u^11_21
                 ... u^n_(n-1) u^n_(n-2) ... u^3_2 u^3_1 u^2_1 |ket>

        where ^ indicates creation operators (a, b; rows) and _ indicates
        annihilation operators (i, j; columns). The doubles amplitudes are
        arbitrarily chosen in the upper-triangular space (a,b <= i,j), but the
        lower-triangular space is used for the individual double pairs
        (a > b, i > j) and for the singles amplitudes (a > i). In all cases,
        row-major ordering is employed.

        The spin cases of a given set of orbitals a, b, i, j are grouped
        together. For singles, spin-up (a) and spin-down (b) amplitudes are
        constrained to be equal and the spin-up operator is on the right (i.e.,
        is applied first). For doubles, the spin case order is

        u|ket> -> ^bb_bb ^ab_ab ^ab_ba ^ba_ab ^ba_ba ^aa_aa |ket>

        For spatial orbital cases in which the same index appears more than
        once, spin cases that correspond to nilpotent (eg., ^pp_qr ^aa_aa),
        undefined (eg., ^pq_pq ^ab_ab), or redundant (eg., ^pq_pq ^ab_ba)
        operators are omitted.

        Args:
            norb : integer
                Total number of spatial orbitals. (0.5 * #spinorbitals)

        Kwargs:
            t1 : ndarray of shape (norb,norb)
                Amplitudes at which to initialize the singles operators
            t2 : None
                NOT IMPLEMENTED. Amplitudes at which to initialize the doubles
                operators

        Returns:
            uop : object of class FSUCCOperator
                The callable UCCSD operator
    '''
    t1_idx = np.tril_indices (norb, k=-1)
    ab_idxs, ij_idxs = list (t1_idx[0]), list (t1_idx[1])
    pq = [(p, q) for p, q in zip (*np.tril_indices (norb))]
    for ab, ij in combinations_with_replacement (pq, 2):
        ab_idxs.append (ab)
        ij_idxs.append (ij)
    uop = FSUCCOperator (norb, ab_idxs, ij_idxs)
    x0 = uop.get_uniq_amps ()
    if t1 is not None: x0[:len (t1_idx[0])] = t1[t1_idx]
    if t2 is not None: raise NotImplementedError ("t2 initialization")
    uop.set_uniq_amps_(x0)
    return uop
\end{lstlisting}

\begin{lstlisting}[caption={\change{Code for generating spin-orbital excitation operators of UCC(SD) in \textit{mrh}. The function ``spincases'' is presented in Listing \ref{lst:spincases}}}, language=Python, float=*, label=lst:spinexc]
class FSUCCOperator (uccsd_sym0.FSUCCOperator):
    ''' A callable spin-adapted (Sz only) unrestricted coupled cluster
        operator. For single-excitation operators, spin-up and spin-down
        amplitudes are constrained to be equal. All spin cases for a given
        spatial-orbital excitation pattern (from i_idxs to a_idxs) are grouped
        together and applied to the ket in ascending order of the index

        (a_spin) * nelec + i_spin

        where 'a_spin' and 'i_spin' are the ordinal indices of the spin
        cases returned by the function 'spincases' for a_idxs (creation
        operators) and i_idxs (annihilation operators) respectively, and nelec
        is the order of the generator (1=singles, 2=doubles, etc.) Nilpotent
        or undefined spin cases (i.e., because of spatial-orbital index
        collisions) are omitted.
        '''


    def __init__(self, norb, a_idxs, i_idxs):
        # Up to two equal indices in one generator are allowed
        # However, we still can't have any equal generators
        self.a_idxs = []
        self.i_idxs = []
        self.symtab = []
        for ix, (a, i) in enumerate (zip (a_idxs, i_idxs)):
            a = np.ascontiguousarray (a, dtype=np.uint8)
            i = np.ascontiguousarray (i, dtype=np.uint8)
            errstr = 'a,i={},{} invalid for number-sym op'.format (a,i)
            assert (len (a) == len (i)), errstr
            #errstr = 'a,i={},{} degree of freedom undefined'.format (a,i)
            #assert (not (np.all (a == i))), errstr
            if len (a) == 1: # Only case where I know the proper symmetry
                             # relation between amps to ensure S**2
                symrow = [len (self.a_idxs), len (self.i_idxs)+1]
                self.a_idxs.extend ([a, a+norb])
                self.i_idxs.extend ([i, i+norb])
                self.symtab.append (symrow)
            else:
                for ix_ab, (ab, ma) in enumerate (zip (*spincases (a, norb))):
                    if np.amax (np.unique (ab, # nilpotent escape
                        return_counts=True)[1]) > 1: continue
                    for ix_ij, (ij, mi) in enumerate (zip (*spincases (
                            i, norb))):
                        if mi != ma: continue # sz-break escape
                        if np.all (ab==ij): continue # undefined escape
                        if np.all (a==i) and ix_ab>ix_ij:
                            continue # redundant escape
                        if np.amax (np.unique (ij, return_counts=True)[1]) > 1:
                            continue # nilpotent escape
                        self.symtab.append ([len (self.a_idxs)])
                        self.a_idxs.append (ab)
                        self.i_idxs.append (ij)
        self.norb = 2*norb
        self.ngen = len (self.a_idxs)
        assert (len (self.i_idxs) == self.ngen)
        self.uniq_gen_idx = np.array ([x[0] for x in self.symtab])
        self.amps = np.zeros (self.ngen)
        self.assert_sanity ()
\end{lstlisting}

\begin{lstlisting}[caption={\change{Code for generating a valid list of spin cases corresponding to a set of field operators for particular spatial orbitals in \textit{mrh}}}, language=Python, float=*, label=lst:spincases]
def spincases (p_idxs, norb):
    ''' Compute the spinorbital indices corresponding to all spin cases of a
        set of field operators acting on a specified list of spatial orbitals
        The different spin cases are returned 'column-major order':

        aaa...
        baa...
        aba...
        bba...
        aab...

        The index of a given spincase string ('aba...', etc.) can be computed
        as

        p_spin = int (spincase[::-1].replace ('a','0').replace ('b','1'), 2)

        Args:
            p_idxs : ndarray of shape (nelec,)
                Spatial orbital indices
            norb : integer
                Total number of spatial orbitals

        Returns:
            p_idxs : ndarray of shape (2^nelec, nelec)
                Rows contain different spinorbital cases of the input spatial
                orbitals
            m : ndarray of shape (2^nelec,)
                Number of beta (spin-down) orbitals in each spin case
    '''
    nelec = len (p_idxs)
    p_idxs = p_idxs[None,:]
    m = np.array ([0])
    for ielec in range (nelec):
        q_idxs = p_idxs.copy ()
        q_idxs[:,ielec] += norb
        p_idxs = np.append (p_idxs, q_idxs, axis=0)
        m = np.append (m, m+1)
    p_sorted = np.stack ([np.sort (prow) for prow in p_idxs], axis=0)
    idx_uniq = np.unique (p_sorted, return_index=True, axis=0)[1]
    p_idxs = p_idxs[idx_uniq]
    m = m[idx_uniq]
    return p_idxs, m
\end{lstlisting}
\change{
For the UCCSD runs of the main text, Fig.~(4), we use ordering
consistent with Qiskit's UCCSD generation. 
We have included a short code listing (Listing~\ref{lst:hfuccsdexc}) which will reproduce the excitation ordering.
}
\begin{lstlisting}[caption={\change{Code for generating the excitation operators of the UCCSD ansatz for the UCCSD calculations of Fig.~(4) of the main text. }}, language=Python, float=*, label=lst:hfuccsdexc]
def generate_fermionic_excitations_occ(num_excitations,num_spin_orbitals,num_particles,
                                   alpha_occ,alpha_unocc,beta_occ,beta_unocc):
    #Assumes ordering [occ_alpha,unocc_alpha,occ_beta,unocc_beta]
    alpha_excitations = []
    # generate alpha-spin orbital indices for occupied and unoccupied ones
    # the Cartesian product of these lists gives all possible single alpha-spin excitations
    alpha_excitations = list(itertools.product(alpha_occ, alpha_unocc))

    # the Cartesian product of these lists gives all possible single beta-spin excitations
    beta_excitations = list(itertools.product(beta_occ, beta_unocc))

    # we can find the actual list of excitations by doing the following:
    #   1. combine the single alpha- and beta-spin excitations
    #   2. find all possible combinations of length `num_excitations`
    pool = itertools.combinations(
        alpha_excitations + beta_excitations, num_excitations
    )

    excitations = list()
    visited_excitations = set()

    for exc in pool:
        # validate an excitation by asserting that all indices are unique:
        #   1. get the frozen set of indices in the excitation
        exc_set = frozenset(itertools.chain.from_iterable(exc))
        #   2. all indices must be unique (size of set equals 2 * num_excitations)
        #   3. and we palso don't want to include permuted variants of identical excitations
        if len(exc_set) == num_excitations * 2 and exc_set not in visited_excitations:
            visited_excitations.add(exc_set)
            occ, unocc = zip(*exc)
            exc_tuple = (occ, unocc)
            excitations.append(exc_tuple)

    return excitations

\end{lstlisting}

\bibliography{library.bib}